\let\bibliographystyle\relax

\documentclass[%
 reprint,
superscriptaddress,
 amsmath,amssymb,
 aps,
prb,
]{revtex4-2}

\usepackage{graphicx}
\usepackage{dcolumn}
\usepackage{bm}
\usepackage[utf8]{inputenc}
\usepackage[T1]{fontenc}
\usepackage{siunitx}
\usepackage{nameref}
\usepackage{amsmath}
\usepackage{amstext}
\usepackage{mathtools} 
\usepackage{amsmath,amssymb}    
\usepackage{booktabs}
\usepackage{array}
\usepackage{graphicx}
\usepackage{xcolor}
\usepackage{placeins}
\usepackage{tabularx}
\usepackage{dblfloatfix}
\usepackage{blindtext}
\usepackage{lipsum}
\usepackage{natbib}
\usepackage{hyperref}
\usepackage{floatrow}
\usepackage{microtype}
\usepackage[caption=false]{subfig}
\usepackage{booktabs}
\DeclareSIUnit{\rad}{rad}

\hypersetup{
        colorlinks   = true,
        citecolor    = blue,
        linkcolor    = blue}

\DeclareSIUnit\torr{Torr}

\begin{document}
%\preprint{APS/123-QED}

\title{Engineering the spin-orbit torque efficiency and magnetic properties of Tb/Co ferrimagnetic multilayers by stacking order}

\author{Mickey Martini}
\affiliation{Department of Materials, ETH Zürich, Hönggerbergring 64 CH-8093 Zürich, Switzerland}
\affiliation{Department of Physics, Politecnico di Milano, P.za Leonardo da Vinci 32, I-20133 Milano, Italy}
\author{Can Onur Avci}
\affiliation{Department of Materials, ETH Zürich, Hönggerbergring 64 CH-8093 Zürich, Switzerland}
\author{Silvia Tacchi}
\affiliation{Istituto Officina dei Materiali del CNR (CNR-IOM), Sede Secondaria di Perugia, c/o Dipartimento di Fisica e Geologia, Università di Perugia, I-06123 Perugia, Italy}
\author{Charles-Henri Lambert}
\affiliation{Department of Materials, ETH Zürich, Hönggerbergring 64 CH-8093 Zürich, Switzerland}
\author{Pietro Gambardella}
\affiliation{Department of Materials, ETH Zürich, Hönggerbergring 64 CH-8093 Zürich, Switzerland}

\received{\today}

%Abstract             
\begin{abstract}

We measured the spin-orbit torques (SOTs), current-induced switching, and domain wall (DW) motion in synthetic ferrimagnets consisting of Co/Tb layers with differing stacking order grown on a Pt underlayer. We find that the SOTs, magnetic anisotropy, compensation temperature and SOT-induced switching are highly sensitive to the stacking order of Co and Tb and to the element in contact with Pt. Our study further shows that Tb is an efficient SOT generator when in contact with Co, such that its position in the stack can be adjusted to generate torques additive to those generated by Pt. With optimal stacking and layer thickness, the dampinglike SOT efficiency reaches up to 0.3, which is more than twice that expected from the Pt/Co bilayer. Moreover, the magnetization can be easily switched by the injection of pulses with current density of about $0.5{-}2  {\times}10^7\SI{}{\ampere\per\centi\meter^2}$ despite the extremely high perpendicular magnetic anisotropy barrier (up to 7.8 T). Efficient switching is due to the combination of large SOTs and low saturation magnetization owing to the ferrimagnetic character of the multilayers. We observed current-driven DW motion in the absence of any external field, which is indicative of homochiral Néel-type DWs stabilized by the interfacial Dzyaloshinkii-Moriya interaction. These results show that the stacking order in transition metal/rare-earth synthetic ferrimagnets plays a major role in determining the magnetotransport properties relevant for spintronic applications.

\end{abstract}
\pacs{Valid PACS appear here}% PACS, the Physics and Astronomy
                             % Classification Scheme.

\maketitle

%Introduction
\section{\label{sec:level1}Introduction}
Magnetization switching and domain wall (DW) motion driven by spin-orbit torques (SOTs) have attracted extensive research interest due to their potential applications in nonvolatile memory and logic technologies \cite{miron2011perpendicular,liu2012spin,garello2014ultrafast,oh2016field,fukami2016spin,avci2017current,grimaldi2020single,miron2011fast,emori2013current,ryu2013chiral,avci2019interface,luo2020current,manchon2019current,dieny2020opportunities}. To improve the scalability and thermal stability of devices in these applications, materials with perpendicular magnetic anisotropy (PMA) are often preferred. Ultrathin films of $3d$ transition metals and alloys (e.g., Co or CoFeB) have been widely investigated for such purposes, where the interfacial spin-orbit coupling gives rise to the PMA \cite{carcia1988perpendicular,gambardella2003giant,ikeda2010perpendicular,dieny2017perpendicular}. Recently, however, increasing interest has been devoted to ferrimagnetic alloys of rare earth metals and transition metals possessing bulk PMA \cite{zhao2015spin,roschewsky2016spin,ueda2016spin,ueda2017temperature,finley2016spin,roschewsky2017spin, mishra2017anomalous,seung2017temperature,siddiqui2018current,pham2018thermal,je2018spin,zheng2019enhanced,caretta2018fast,cai2020ultrafast,sala2021real,ueda2016effect,bang2016enhancement,yu2019long,wong2019enhanced}. In these compounds, the rare-earth-metal and transition-metal magnetic sublattices are coupled antiparallel to each other, resulting in a reduced total magnetic moment relative to the individual sublattices \cite{buschow1977intermetallic, hansen1989magnetic}. The low magnetization allows for efficient SOT-induced switching, whereas the large PMA increases the thermal stability of the devices. Moreover, the possibility of obtaining angular momentum compensation allows for antiferromagnetlike magnetization dynamics that, e.g., leads to ultrafast switching and DW motion \cite{caretta2018fast, cai2020ultrafast, sala2021real}. 
To date, studies of SOTs and related phenomena have mainly focused on the alloy form of the rare-earth-metal–transition-metal systems \cite{zhao2015spin,roschewsky2016spin,ueda2016spin,ueda2017temperature,finley2016spin,roschewsky2017spin, mishra2017anomalous,seung2017temperature,siddiqui2018current,pham2018thermal,je2018spin,zheng2019enhanced,caretta2018fast,cai2020ultrafast,sala2021real}, whereas multilayer structures have been less explored \cite{ueda2016effect,bang2016enhancement,yu2019long,wong2019enhanced}. In such structures, the stacking order and the thickness of each layer offer additional degrees of freedom to tune the SOTs and understand the role of each element in interface-driven phenomena \cite{manchon2019current}. Theoretical \cite{tanaka2010intrinsic} and experimental \cite{ueda2016effect,reynolds2017spin,wong2019enhanced,wu2020spin,cespedes2021current,krishnia2021spin} studies suggest that the rare-earth elements with partially filled {\it f} orbitals are candidates for generating large spin Hall effect, and hence can be used as a source of SOTs. The strong spin-orbit interaction in the rare earth metals combined with the inversion symmetry breaking at their interfaces with transition metals can also produce sizeable Rashba-Edelstein effect and Dzyaloshinkii-Moriya interaction (DMI). The latter stabilizes homochiral Néel-type domain walls with efficient and ultrafast current-induced dynamics \cite{emori2013current,ryu2013chiral,avci2019interface,thiaville2012dynamics,martinez2014current,baumgartner2017spatially}.

Motivated by the above considerations, we investigate the influence of the stacking order and layer thickness on current-induced SOTs, switching, and DW motion in Pt/Co/Tb-based multilayers. Harmonic Hall and magneto-optic Kerr effect measurements reveal strong SOTs in the perpendicularly magnetized Tb/Co multilayers, which lead to efficient current-induced switching and DW motion, comparable to those of TbCo alloys in contact with Pt. Moreover, the SOTs and other magnetic properties, such as the saturation magnetization, magnetic anisotropy, compensation temperature, and the anomalous Hall effect (AHE) are highly sensitive to the stacking order as well as to the element in contact with the Pt underlayer. We find that Tb can be an effective source of SOTs with additive contributions to the ones due to Pt when placed on top of Pt/Co. These results show that the stacking order in rare-earth-metal/transition-metal multilayers offers an additional degree of freedom to tune the efficiency of SOT-controlled switching and DW motion. 

%preparation
\FloatBarrier
\section{Sample preparation and characterization}\label{Section:preparation}

 \begin{figure}[b!]
    \centering
    \includegraphics[width=1\textwidth]{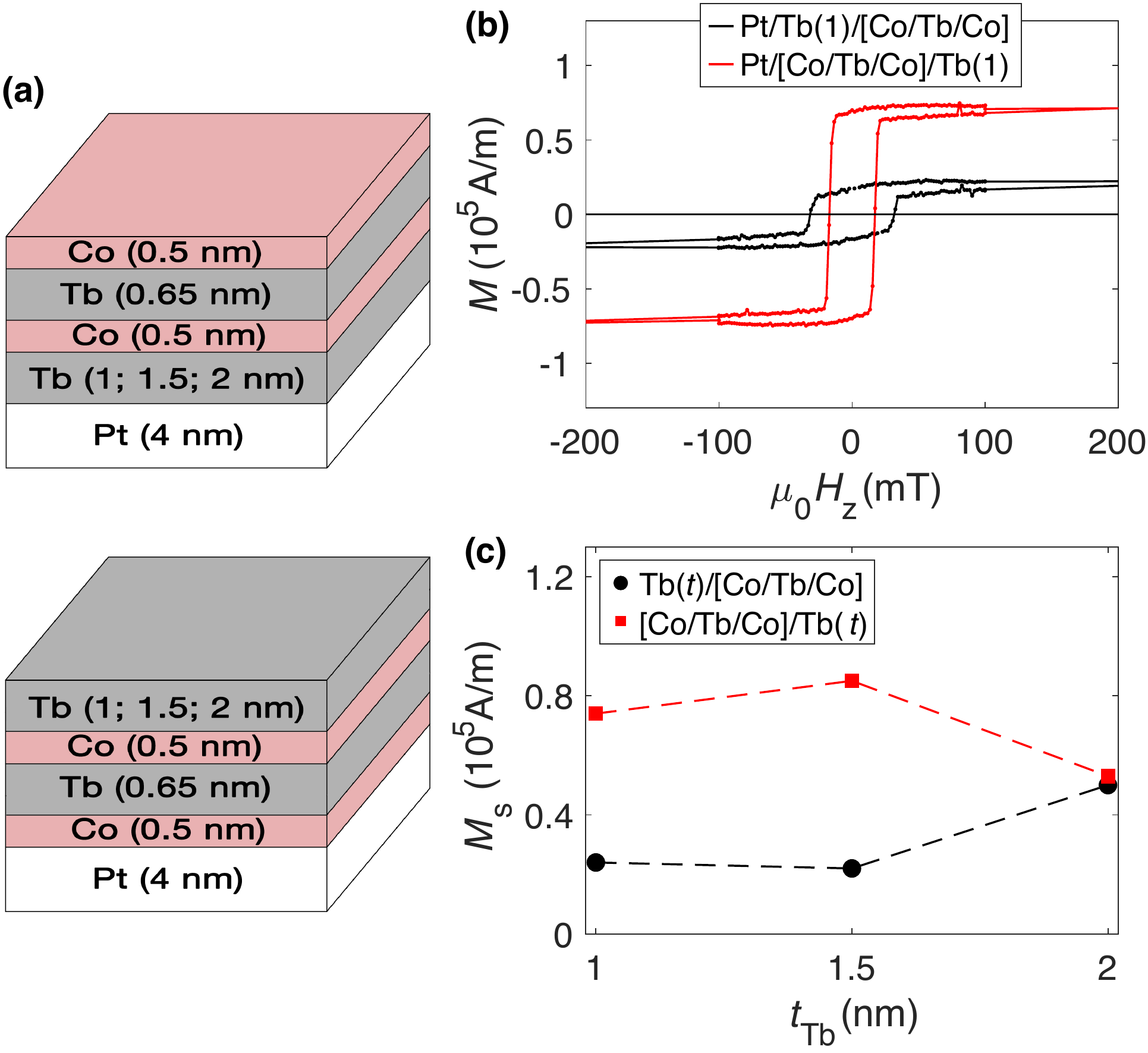}
    \caption{(a) Schematic illustration of the Co/Tb/Co multilayers with different stacking order. (b) Magnetization curves of Pt/Tb(1)/[Co/Tb/Co] and Pt/[Co/Tb/Co]/Tb(1). (c) Saturation magnetization versus Tb thickness.}
    \label{fig:charact}
\end{figure}

We grow Pt(4)/Tb($t_\mathrm{Tb}$)/\textbf{[}Co(0.5)/Tb(0.65)/Co(0.5)\textbf{]} and Pt(4)/\textbf{[}Co(0.5)/Tb(0.65)/Co(0.5)\textbf{]}/Tb($t_\mathrm{Tb}$) multilayers with PMA as well as Pt(5)/Tb($t_\mathrm{Tb}$)/Co(5) and  Pt(5)/Co(5)/Tb($t_\mathrm{Tb}$) trilayers with in-plane anisotropy on thermally oxidized silicon substrates by dc magnetron sputtering (numbers indicate the thickness in nm). For all layers, a Ti capping of $\SI{5}{\nano\meter}$ and a Ta seed layer of \SI{3}{\nano\meter} are used. The base pressure and Ar partial pressure of the deposition chamber are $\SI{6e-8}{\milli\bar}$ and $\SI{4 e-3}{\milli\bar}$, respectively. The first two series, schematically illustrated in Fig.~\hyperlink{fig:charact}{1(a)}, are used for the SOTs, current-induced switching, and DW motion studies, whereas in the last two series only the SOTs are quantified. In the samples with PMA we fix the layers between square brackets and change the Tb thickness ($t_\mathrm{Tb}$) and position relative to them; similarly, in the trilayers with in-plane anisotropy we changed the Tb thickness and position relative to Co. Hall bar devices are fabricated by patterning the substrates using optical lithography before deposition and subsequent lift-off. 

In Fig.~\hyperref[fig:charact]{1(b)} we show representative magnetization curves measured by a superconducting quantum interference device at room temperature with an out-of-plane field sweep for the two different positions of the Tb(1) layer with respect to the fixed [Co/Tb/Co] trilayer. Both sets of data show clear hysteresis and nearly 100$\%$ remanence at zero field, indicative of strong PMA. However, their saturation magnetization ($M$\textsubscript{s}) differs significantly even though the amount of Tb and Co is the same in both multilayers. Measurements of all the six samples with PMA [Figure~\hyperref[fig:charact]{1(c)}] show that $M$\textsubscript{s} is larger when Co is in direct contact with Pt except for the thickest Tb samples. In both multilayer systems, $M_\mathrm{s}$ is significantly smaller relative to that expected for Co (>10$^6$ A/m), as expected due to the antiferromagnetic coupling of Co and Tb. The large difference in $M$\textsubscript{s} and its dependence on Tb thickness indicate that the net magnetization is highly sensitive to the stacking order. 

Figure \hyperlink{fig:compensation}{2(a)} shows the device layout with the electrical connections and the coordinate system. Figure~\hyperlink{fig:compensation}{2(b)} reports the sheet resistance $R_\mathrm{s}$ measured using a four-point geometry in the structures with PMA. $R_\mathrm{s}$ is about 20$\%$ larger in samples with Pt/Tb interface compared with those with Pt/Co interface. The larger $R_\mathrm{s}$ and its increase with $t_\mathrm{Tb}$ in Pt/Tb($t_\mathrm{Tb}$)/[Co/Tb/Co] might be due to increased interfacial electron scattering and/or intermixing between Tb and Pt in this system. In contrast, in Pt/[Co/Tb/Co]/Tb($t_\mathrm{Tb}$), $R_\mathrm{s}$ does not significantly change with the Tb thickness, suggesting that the electric conductivity is predominantly dictated by the Pt and Co layers. The resistivity $\rho$ of all six multilayers with PMA, estimated by considering only the conduction through Pt, Co, and Tb and neglecting the contribution from the buffer and capping layers is reported in Table \ref{tab:table} along with the other properties measured in this study. 

\begin{figure}[h!]
\centering
\includegraphics[width=1\textwidth]{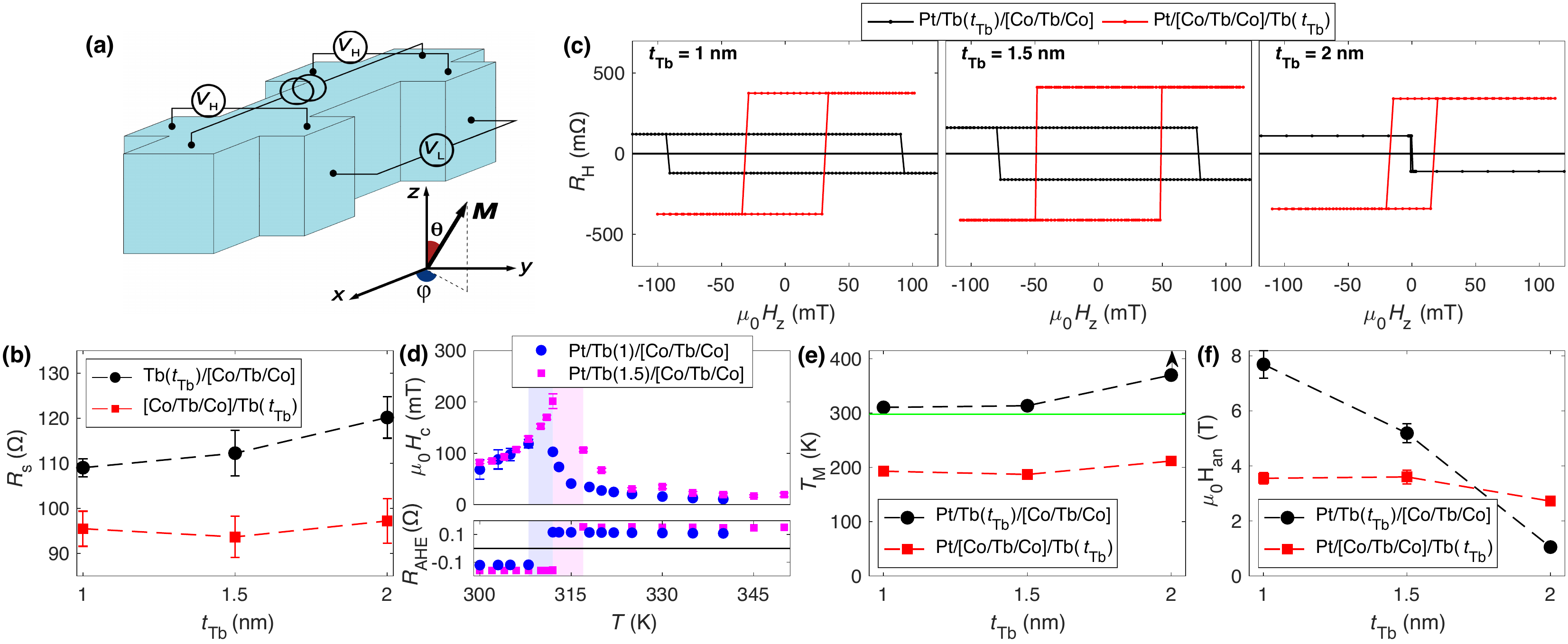}
\caption{(a) Schematics of a Hall bar device and coordinate system.  The Hall bar channel is $\SI{100}{\micro\meter}$ long and $\SI{5}{\micro\meter}$ wide. (b) Sheet resistance versus thickness of Tb layer in Pt/Tb($t_\mathrm{Tb}$)/[Co/Tb/Co] and Pt/[Co/Tb/Co]/Tb($t_\mathrm{Tb}$). (c) Hall resistance as a function of out-of-plane magnetic field in samples with different Tb thickness. (d) Coercive field versus temperature in Pt/Tb(1,1.5)/[Co/Tb/Co]. The regions shaded in blue and magenta represent temperature ranges in which the magnetization is strongly compensated. (e) Magnetization compensation temperature as a function of Tb thickness. The black arrow indicates that $T_M> \SI{370}{\kelvin}$, which is the largest temperature reached in our setup; the green solid line indicates room temperature. (f) Anisotropy field at room temperature versus Tb thickness.}
\label{fig:compensation}
\end{figure}

In alloys of rare earth metals and transition metals and in multilayers, the magnetotransport properties are dominated by the 3\textit{d}4\textit{s} orbitals of the transition-metal elements near the Fermi level; thus the AHE is mainly driven by the transition-metal sublattice \cite{mimura1976hall, ueda2016effect, roschewsky2016spin}. In Fig.~\hyperref[fig:compensation]{2(c)} we report the Hall resistance $R$\textsubscript{H} as a function of out-of-plane magnetic field. As evident from the data, the AHE is positive and large in the Pt/[Co/Tb/Co]/Tb series as opposed to the Pt/Tb/[Co/Tb/Co] series, where the AHE is negative and relatively small. This means that \textit{i}) at room temperature the total magnetization is dominated by the Co (Tb) sublattice for the layers with Pt/Co (Pt/Tb) interface and that the magnetic compensation temperature $T_\mathrm{M}$ is below (above) room temperature; \textit{ii}) the Pt/Co interface has a significant contribution to the AHE, in agreement with literature data \cite{avci2019effects}. Our data clearly show that the stacking order plays a crucial role in determining the magnetic compensation point as well as the AHE, in addition to $M$\textsubscript{s} and $R_\mathrm{s}$ discussed earlier. We further study the coercive field ($H$\textsubscript{c}) as a function of temperature to characterize $T_\mathrm{M}$ of all six samples. As the temperature crosses $T_\mathrm{M}$, the anomalous Hall resistance $R$\textsubscript{H} changes sign; further, $H$\textsubscript{c} diverges because a stronger external field is required to reverse the vanishing net magnetization \cite{webb1988coercivity}, as shown for Pt/Tb(1; 1.5~nm)/[Co/Tb/Co] in Fig.~\hyperref[fig:compensation]{2(d)}. $T_\mathrm{M}$ determined from the change of sign of $R$\textsubscript{H} is shown in Fig.~\hyperref[fig:compensation]{2(e)} for all the six samples. The stacking order gives rise to a substantial difference of more than 100 K in $T_\mathrm{M}$ between Pt/Tb($t_\mathrm{Tb}$)/[Co/Tb/Co] and Pt/[Co/Tb/Co]/Tb($t_\mathrm{Tb}$), which is likely related to the larger $M$\textsubscript{s} of Pt/[Co/Tb/Co]/Tb($t_\mathrm{Tb}$).

Finally, we characterize the effective magnetic ansitropy field $H_\mathrm{an}$ by measuring $R_\mathrm{H}$ in the samples with PMA as a function of an external magnetic field applied nearly in plane at polar and azimuthal angles $\theta_\mathrm{H}=\SI{88}{\degree}$ and $\varphi_\mathrm{H} = \SI{90}{\degree}$ [see Fig.~\hyperlink{fig:compensation}{2(a)} for the definition of the angular coordinates]. We use a macrospin approximation to calculate $H_\mathrm{an}= H\left(\frac{\sin{\theta_H}}{\sin{\theta}}-\frac{\cos{\theta_H}}{\cos{\theta}}\right)$, where $\theta = \arccos{\left|R_\mathrm{H}\big/R_\mathrm{AHE}\right|}$ \cite{avci2014fieldlike}, as reported in Fig.~\hyperlink{fig:compensation}{2(f)}. We find $\mu_0 H_\mathrm{an}=7.8\pm0.5$~T in Pt/Tb(1)/[Co/Tb/Co], which rapidly decreases with increasing Tb thickness due to the shift of $T_\mathrm{M}$ to higher temperatures. In the  Pt/[Co/Tb/Co]/Tb(1) series, on the other hand, $H_\mathrm{an}$ is relatively constant, which is consistent with the behavior of $T_\mathrm{M}$ in these samples. We note that extremely large anisotropy fields in excess of 3~T are measured in all of our samples, with the exception of Pt/Tb(2)/[Co/Tb/Co].  

\onecolumngrid

\twocolumngrid

%Results
%SOTs
\section{Results and Discussion}

\subsection{SOTs in multilayers with PMA}\label{Sect:SOTs}

The SOT-induced effective fields are quantified by performing harmonic Hall effect measurements in the PMA samples \cite{garello2013symmetry, ghosh2017interface}. An alternating current with frequency $\omega/2\pi = \SI{10}{\hertz}$  and amplitude $j =0.3{-}1{\times}10^7 \SI{}{\ampere\per\centi\meter^2}$ is applied along the \textit{x} axis, and the first- and second-harmonic Hall resistances ($R_\mathrm{H}^\mathrm{1\omega}$ and $R_\mathrm{H}^\mathrm{2\omega}$) were measured during an in-plane field sweep along the \textit{x} and \textit{y} axes ($H_\mathrm{x}, H_\mathrm{y}$). 
The current density is estimated by dividing the current by the total cross section of the Pt, Co, and Tb layers. We neglect the conductivity of the highly resistive Ta buffer and Ti capping layers. In multilayers with inversion symmetry along the $z$ axis and current injection along $x$, the dampinglike and fieldlike SOT effective fields are given by $\mathbf{H}_\mathrm{DL}= H_\mathrm{DL} (\mathbf{y}\times\hat{\mathbf{M}})$ and $\mathbf{H}_\mathrm{FL}= H_\mathrm{FL} \mathbf{y}$, respectively, where $\mathbf{M}$ is the unit vector of the magnetization and the current-induced spin accumulation is polarized along $\mathbf{y}$~\cite{manchon2019current}. In order to define the sign of the field amplitudes $H_\mathrm{DL,FL}$, we adopt the convention that $H_\mathrm{DL}$ ($H_\mathrm{FL}$) is positive (negative) when the field points in the same direction as in Pt/Co (opposite direction to the Oersted field), where Pt is deposited before Co \cite{garello2013symmetry}. 
Due to the large PMA in our samples, we opt for the small-angle approximation method where the magnetization is slightly tilted away from equilibrium by the applied field. In this regime, the effective fields can be estimated using the following relations \cite{kim2013layer, hayashi2014quantitative}:

\begin{align}
    \Delta H_\mathrm{x(y)} &= \left(\frac{\partial R_\mathrm{H}^\mathrm{2\omega}}{\partial H_\mathrm{x(y)}}\bigg/ \frac{\partial^2 R_\mathrm{H}^\mathrm{1\omega}}{\partial H_\mathrm{x(y)}^2}\right),\label{eq:dH}\\
      H_\mathrm{DL(FL)} &= -2 \frac{\Delta H_\mathrm{x(y)} \pm \nu \Delta H_\mathrm{y(x)}}{1-4\nu^2}\label{eq:Heff},
\end{align}

%Axes labels too big? Please try to use img/torque (without 2)

\noindent
where $\nu$ is the ratio of the planar Hall resistance $R$\textsubscript{PHE} to the anomalous Hall resistance $R$\textsubscript{AHE} and $\pm$ corresponds to the measurements taken when the magnetization is pointing along the $\pm z$ direction. To characterize $R$\textsubscript{AHE} and $R$\textsubscript{PHE}, we use the first-harmonic signal given by $R_\mathrm{H}^\mathrm{\omega} = R_\mathrm{AHE}\cos{\theta} + R_\mathrm{PHE}\sin^2{\theta}\sin{2\varphi}$. Since the magnetization cannot be fully saturated in plane by the available field, we quantify $R$\textsubscript{PHE} by measuring $R_\mathrm{H}^\mathrm{\omega}$ with an external field applied at $(\theta_\mathrm{H},\varphi_\mathrm{H})\equiv (\SI{88}{\degree},\SI{45}{\degree})$. We then separate the antisymmetric and symmetric signal contributions (with respect to the field reversal), which originate from the first and second terms in the above equation, respectively. The antisymmetric signal is then plotted as a function of $\sin^2{\theta}$, where $\theta$ is estimated using the macrospin approximation as $\theta = \arccos{\left|R_\mathrm{H}^\mathrm{1\omega}\big/R_\mathrm{AHE}\right|}$. Finally the slope of the linear fit represents $R$\textsubscript{PHE}. The details of this approach can be found in Ref.~\cite{garello2013symmetry}. 

\begin{figure}[h!] %b!
\centering
\includegraphics[width=0.7\textwidth]{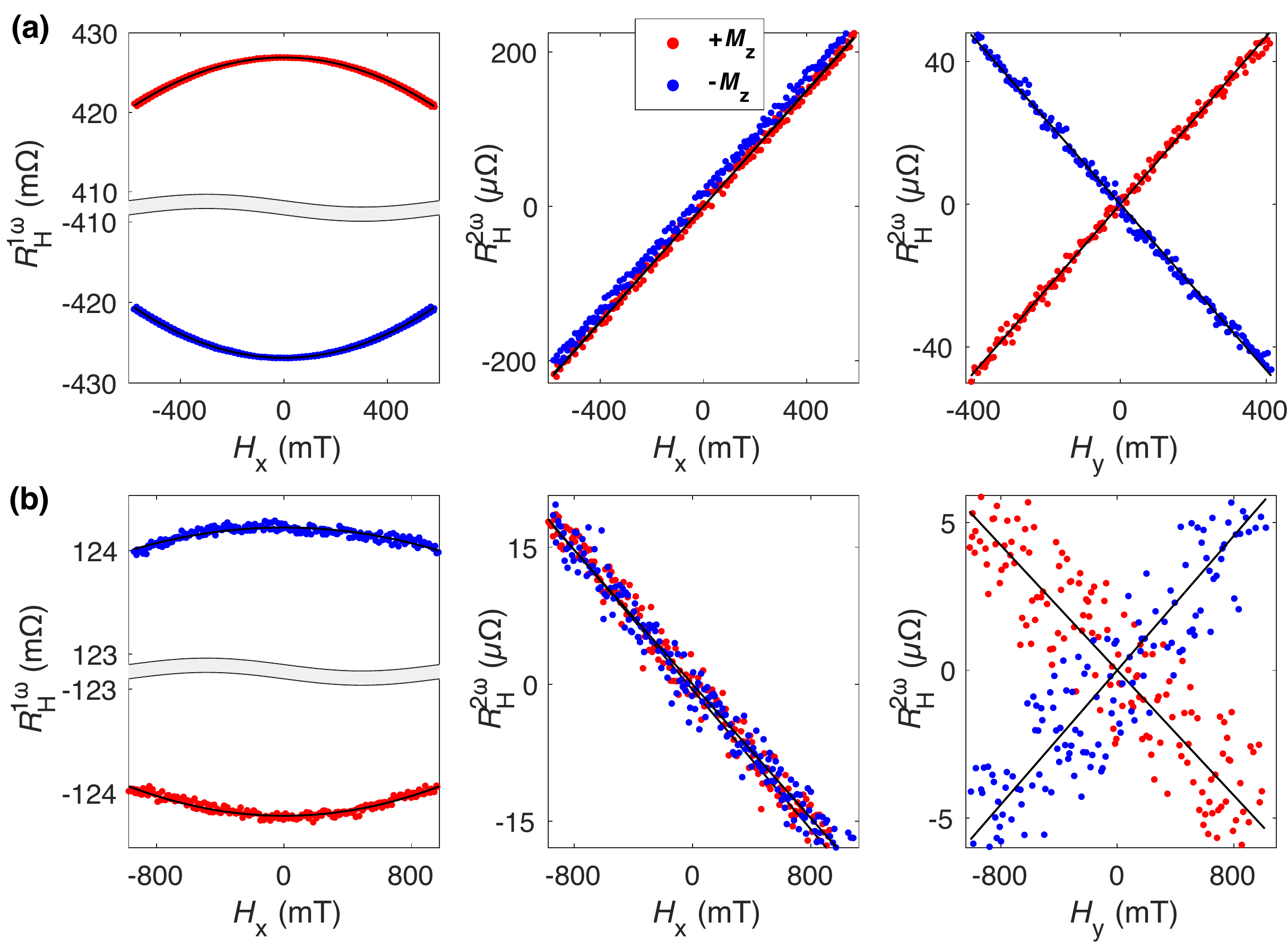}
\caption{First and second harmonic of the Hall resistance ($R_\mathrm{H}^\mathrm{1\omega}$ and $R_\mathrm{H}^\mathrm{2\omega}$, respectively) versus in-plane longitudinal and transverse magnetic fields ($H_\mathrm{x}$ and $H_\mathrm{y}$) measured using $j = \SI{.8e7}{\ampere\per\meter^2}$ in (a) Pt/[Co/Tb/Co]/Tb(1.5) and (b) Pt/Tb(1)/[Co/Tb/Co].}
\label{fig:methods}
\end{figure}

Figure~\ref{fig:methods} shows representative $R_\mathrm{H}^\mathrm{1\omega}$ and $R_\mathrm{H}^\mathrm{2\omega}$ measurements taken from the Pt/[Co/Tb/Co]/Tb(1.5) and Pt/Tb(1)/[Co/Tb/Co] samples. $R_\mathrm{H}^\mathrm{1\omega}$ has a parabolic field dependence due to the coherent rotation of the magnetization. Note that we only plot the data corresponding to the $H_\mathrm{x}$ sweep since the $H_\mathrm{y}$ sweep shows an identical behavior. The sign of $R_\mathrm{H}^\mathrm{\omega}$ is opposite in these two samples for a given magnetization orientation due to the opposite AHE [see Fig.\hyperlink{fig:compensation}{2(c)}]. $R_\mathrm{H}^\mathrm{2\omega}$ varies linearly with both $H$\textsubscript{x} and $H$\textsubscript{y} as expected from the action of SOTs. We note that the slopes of $R_\mathrm{H}^\mathrm{2\omega}$ are opposite in Co-like and Tb-like samples. Considering also the opposite sign of $R_\mathrm{H}^\mathrm{1\omega}$ in these two samples we conclude that the SOT effective fields have the same sign in both systems. Thus, the SOTs act on the net magnetization, in agreement with previous works \cite{ueda2016spin,finley2016spin}. The SOT effective fields normalized by the current density, $\chi_\mathrm{DL(FL)} = \mu_0 H_\mathrm{DL(FL)}\big/j$, are summarized in~Figs.~\hyperlink{torques}{4(a)}-\hyperlink{torques}{(4b)} as a function of Tb thickness. The values of $\chi_\mathrm{DL(FL)}$  are obtained by performing a linear fit of $H_\mathrm{DL(FL)}$ versus $j$ for both magnetization orientations (not shown). The fieldlike component includes also the Oersted field contribution due to the current flowing through Pt, which we estimate as $\mu_0 H_\mathrm{Oe}\big/j = \SI{0.3}{\milli\tesla}$ $10^{-7}\SI{}{\ampere^{-1}\centi\meter^2}$ by Ampere's law \cite{avci2014fieldlike}. The Oersted field points in the opposite direction to the fieldlike component of the SOT field, and can be neglected due to its small amplitude compared with $H_\mathrm{FL}$. We observe that $\chi_\mathrm{DL}$ is very strong in Pt/Tb(1)/[Co/Tb/Co], reaching up to $+\SI{72}{\milli\tesla}$ $10^{-7}\SI{}{\ampere^{-1}\centi\meter^2}$ but rapidly decreasing as the Tb thickness is increased. Also $\chi_\mathrm{FL}$ reaches remarkable values, about $+\SI{25}{\milli\tesla}$  $10^{-7}\SI{}{\ampere^{-1}\centi\meter^2}$ in both Pt/Tb(1)/[Co/Tb/Co] and Pt/[Co/Tb/Co]/Tb(1). To compare the SOTs in different layers on an equal footing we calculate the SOT efficiencies  $\xi_\mathrm{DL(FL)}^\mathrm{j} = \frac{2e}{\hbar}M_\mathrm{s}t_\mathrm{M}\chi_\mathrm{DL(FL)}$ \cite{pai2015dependence}, where $e$ is the electron charge, $\hbar$ is the reduced Planck's constant, and $t_\mathrm{M}$ is the magnetic layer thickness. The SOT efficiencies are plotted in Figs.~\hyperlink{torques}{4(c,}-\hyperlink{torques}{ d)} as a function of Tb thickness. We observe that $\xi_\mathrm{DL}^\mathrm{j}$ and $\xi_\mathrm{FL}^\mathrm{j}$ vary between +0.1 and +0.3 and between +0.03 and +0.2, respectively. Whereas $\xi_\mathrm{DL}^\mathrm{j}$ and $\xi_\mathrm{FL}^\mathrm{j}$ in Pt/[Co/Tb/Co]/Tb(1) and Pt/Tb(1)/[Co/Tb/Co] are comparable to those measured in Pt/Co bilayers~\cite{garello2013symmetry,manchon2019current}, we observe that both efficiencies increase strongly as a function of Tb thickness in  Pt/[Co/Tb/Co]/Tb($t$), reaching values that are about two times larger compared with Pt/Co. Moreover, $\xi_\mathrm{FL}^\mathrm{j}$ is significantly larger in Pt/[Co/Tb/Co]/Tb($t_\mathrm{Tb}$) compared with that in Pt/Tb($t_{\mathrm{Tb}}$)/[Co/Tb/Co], and has a different thickness dependence in the two series. Similar trends are also observed if we consider the SOT efficiency per unit electric field $\xi_\mathrm{DL(FL)}^\mathrm{E} = \xi_\mathrm{DL(FL)}^\mathrm{j}\big/\rho$ (see Table \ref{tab:table}).     

\onecolumngrid

\twocolumngrid

\onecolumngrid
\newpage
\begin{figure}[h!] %t
\includegraphics[width=.84\textwidth]{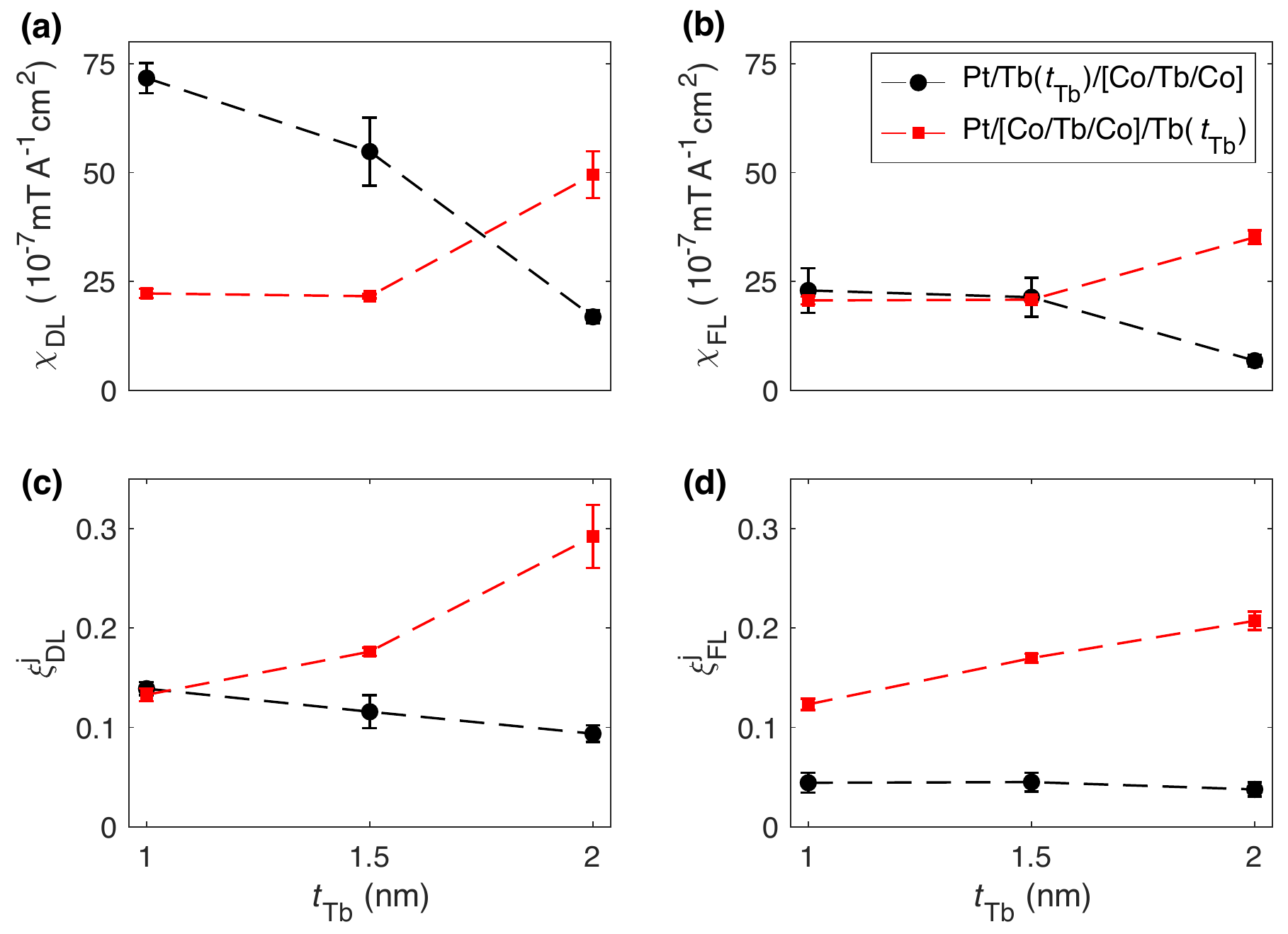}
{\caption{(a) Dampinglike and (b) fieldlike SOT effective field per unit current density $\chi_\mathrm{DL,FL}$ versus Tb thickness in Pt/[Co/Tb/Co]/Tb($t_{\mathrm{Tb}}$) and (b) Pt/Tb($t_{\mathrm{Tb}}$)/[Co/Tb/Co]. (c) Dampinglike and (d) fieldlike SOT efficiency $\xi_\mathrm{DL,FL}$ versus Tb thickness. The error bars take into account the errors of the linear and parabolic fits of $R_\mathrm{H}^\mathrm{1\omega}$ and $R_\mathrm{H}^\mathrm{2\omega}$ respectively, averaged according to Eq.~\ref{eq:Heff} for each current density, propagated in the linear fits of $H_\mathrm{DL(FL)}(j)$ for the $\bf +M_\mathrm{z}$ and $\bf -M_\mathrm{z}$ configurations.}\label{fig:torques}}
\end{figure}

 \twocolumngrid
 
 \raggedbottom
 \noindent
  We note that $\xi_\mathrm{DL}^\mathrm{j}$ and $\xi_\mathrm{FL}^\mathrm{j}$ are obtained by normalization to the average current density $j$ flowing in the multilayers, such that even larger efficiencies would be obtained by normalization to the current density flowing in Pt, as often done in the literature. 
  The dependence of $\xi_\mathrm{DL(FL)}^\mathrm{j}$ on the stacking order and Tb thickness can have several origins. First, the structural details of the multilayers and of the interface with Pt generally change with the stacking order, which may ultimately lead to changes of the SOTs. This point is supported by the resistivity measurements, which suggest that Tb deposited on Pt induces intermixing between the two species rather than favoring sharp interfaces. Second, the Tb layer can be a source of dampinglike and fieldlike SOTs in itself, with a contribution opposite to that of Pt. Such a scenario, which agrees with a previous study of Pt/[Co/Ni]/Co/Tb multilayers \cite{wong2019enhanced}, is consistent with the enhancement of $\xi_\mathrm{DL(FL)}^\mathrm{j}$ when Pt and Tb are placed on opposite sides of the [Co/Tb/Co] trilayer. Third, placing the Tb layer on top of the stack might provide an efficient sink for the spin current generated by Pt and the Pt/Co interface \cite{qiu2016enhanced}. The monotonic increase of the dampinglike and fieldlike SOTs in this series is consistent with both the second and third explanation taking into account the finite spin diffusion length of Tb. Overall, even if the exact mechanism giving rise to the enhancement of $\xi_\mathrm{DL}^\mathrm{j}$ and $\xi_\mathrm{FL}^\mathrm{j}$ cannot be singled out with certainty, our data indicate that Tb and the stacking order of Co/Tb multilayers play a crucial role in determining the strength of the SOTs.
 
  \subsection{SOTs in Pt/Tb/Co and Pt/Co/Tb trilayers with in-plane magnetic anisotropy}
 \raggedbottom
 
 We quantify the dampinglike SOT efficiency in samples with in-plane anisotropy following the method outlined in Ref.~\cite{avci2014interplay}.  Injection of a relatively small alternating current ($\omega/2\pi = \SI{10}{\hertz}$, $j =0.3{-}1{\times}10^7 \SI{}{\ampere\per\centi\meter^2}$)  gives rise to periodic oscillations of the magnetization about its equilibrium position. The second-harmonic Hall resistance related to the current-induced fields and the thermoelectric signal due to Joule heating can be written as: 
 
 \begin{multline}\label{eq:SOTs_IP}
     R_\mathrm{H}^\mathrm{2\omega} = \left(R_\mathrm{AHE}\frac{H_\mathrm{DL}}{H_\mathrm{ext}}+I_0\alpha\nabla T\right)\cos{\varphi} +\\
     2R_\mathrm{PHE}(2\cos^3{\varphi}-\cos{\varphi})\frac{H_\mathrm{FL}+H_\mathrm{Oe}}{H_\mathrm{ext}+H_\mathrm{dem}}
 \end{multline}
 
 \newpage
 
 \onecolumngrid

 \begin{figure}[h!]
\includegraphics[width=1\textwidth]{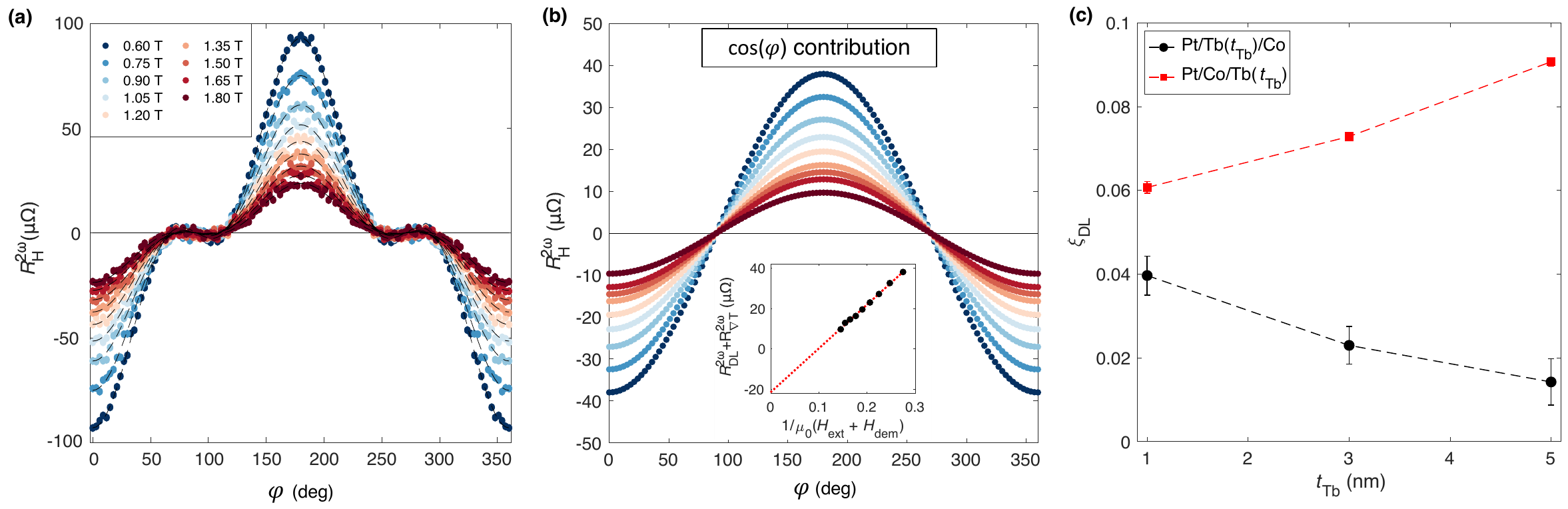}
{\caption{(a) In-plane angular dependence of the second-harmonic Hall resistance of Pt/Co/Tb(3) for $j = {0.8}{\times}10^7 \SI{}{\ampere\per\centi\meter^2}$. (b) Dampinglike and thermal contribution to the second-harmonic Hall resistance. Inset: linear fit of the cosine component of the second-harmonic Hall resistance versus the inverse of the external field and the demagnetizing field acting against the the current-induced field. (c) Dampinglike SOT efficiency versus Tb thickness. The error bars take into account the uncertainty of the linear fit of $H_\mathrm{DL}$ versus $j$.}\label{fig:SOT_IP}}
\end{figure}
 
 \twocolumngrid
 
\begin{table}[b]

\caption{Saturation magnetization, demagnitizing field, dampinglike effective field per unit current density and SOT efficiency in trilayers with in-plane anisotropy. Errors are a fraction of the least significant digit.}
\begin{ruledtabular}
\begin{tabular}{lcccc}
  &$M_\mathrm{s}$ &$\mu_0H_\mathrm{dem}$ &$\chi$\textsubscript{DL} &$\xi_\mathrm{DL}^\mathrm{j}$\\
  &($10^6\SI{}{\ampere\per\meter}$) &$(\SI{}{\tesla}$) &($\SI{}{\milli\tesla}$ $10^{-7}$$\SI{}{\per\ampere}\SI{}{\centi\meter^2}$)&\\
 
\hline
Pt/Tb(1)/Co& 1.13 &1.80 &0.19  &0.040 \\
Pt/Tb(3)/Co& 0.81 &1.59 &0.12 &0.023 \\
Pt/Tb(5)/Co& 0.48 &1.48 &0.10 &0.014 \\
Pt/Co/Tb(1)& 1.22 &0.75 &0.27 &0.061 \\
Pt/Co/Tb(3)& 0.81 &0.72 &0.37 &0.073 \\
Pt/Co/Tb(5)& 0.68 &0.56 &0.44 &0.091 \\

\end{tabular}
\end{ruledtabular}
\label{tab:table_IP}

\end{table} 

\noindent 
where $I_0$ is the amplitude of the injected current, $\alpha$ the anomalous Nernst coefficient, $\nabla T$ the perpendicular thermal gradient and $H_\mathrm{dem}$ the demagnetizing field. Figure \hyperlink{fig:SOT_IP}{5(a)} shows a representative angle scan measurement of $R_\mathrm{H}^\mathrm{2\omega}$ in Pt/Co/Tb(3) for different external in-plane fields along with the fitting curves (dashed lines) given by Eq.~\ref{eq:SOTs_IP}. In Fig.~\hyperlink{fig:SOT_IP}{5(b)} we plot the $\cos{\varphi}$ component of $R_\mathrm{H}^\mathrm{2\omega}$, which includes the combined contributions of dampinglike torque and the Nernst (and/or Seebeck) effect. The amplitude of the $\cos{\varphi}$ component of $R_\mathrm{H}^\mathrm{2\omega}$ decreases on increasing the external field. The latter tends to rigidly align the magnetization along its direction, suppressing the periodic oscillations. This results in a reduction of the SOT and therefore of $R_\mathrm{H}^\mathrm{2\omega}$ and its $\cos{\varphi}$ component. This field dependence can be exploited to further separate the dampinglike torque and the thermoelectric effects in the $\cos{\varphi}$ component of $R_\mathrm{H}^\mathrm{2\omega}$. The anomalous Nernst effect depends only on the magnetization direction and not on the external field amplitude (provided that the magnetization is saturated), whereas the SOT decreases at higher field, as already discussed. Thus, the dampinglike effective field $H_\mathrm{DL}$ can be quantified by performing a linear fit of the $\cos{\varphi}$ component of $R_\mathrm{H}^\mathrm{2\omega}$ versus the inverse of the external field plus the demagnetizing field $H_\mathrm{dem}$~\cite{avci2014interplay}, as shown in the inset of Fig.~\hyperlink{fig:SOT_IP}{5(b)}. This experiment is repeated for all the samples for several current densities to estimate $\chi_\mathrm{DL}$ as the slope of $H_\mathrm{DL}$ versus $j$, as done for the multilayers with PMA in Sect.~\ref{Sect:SOTs}. The values of $\chi_\mathrm{DL}$ are reported in Table \ref{tab:table_IP}. Compared with the multilayers with PMA, the dampinglike effective field in the samples with in-plane anisotropy is 2 orders of magnitude lower. This is due to the large magnetization on which the SOT acts, that is dominated by the bulk magnetic moments of the Co layer (the ferromagnetic layer is now thick). The dampinglike SOT efficiency, calculated also here as $\xi_\mathrm{DL}^\mathrm{j} = \frac{2e}{\hbar}M_\mathrm{s}t_\mathrm{M}\chi_\mathrm{DL}$, is plotted in Fig.~\hyperlink{fig:SOT_IP}{5(c)} as a function of $t_\mathrm{Tb}$. Although $\xi_\mathrm{DL}^\mathrm{j}$ is significantly lower in these samples compared with the samples with PMA, it scales in the same way with $t_\mathrm{Tb}$, increasing (decreasing) in the trilayer with Pt/Co (Pt/Tb) interface. The comparison between Pt/Tb/Co and Pt/Co/Tb structures provides further evidence that the SOTs are very sensitive to $t_\mathrm{Tb}$ and the stacking order of Co/Tb.

 %Switching
 \subsection{Current-induced magnetization switching}
 
 \begin{figure}[b]
    \centering
    \includegraphics[width=.93\textwidth]{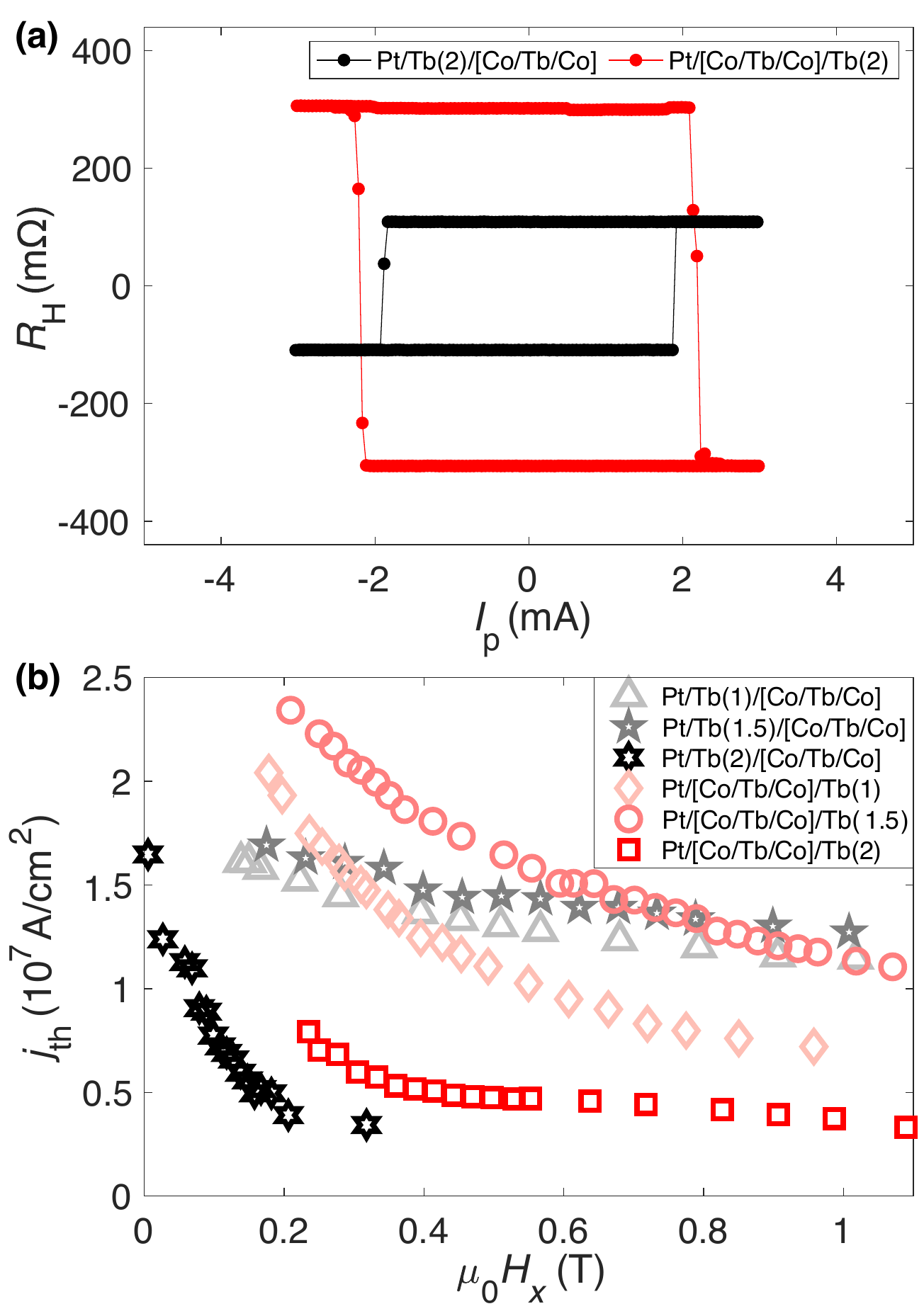}
    \caption{(a) Change of the anomalous Hall resistance due to SOT-induced switching in Pt/Tb(2)/[Co/Tb/Co] with constant in-plane field $\mu_\mathrm{0}H_x = \SI{182}{\milli\tesla}$ and in Pt/[Co/Tb/Co]/Tb(2) with $\mu_\mathrm{0}H_x = \SI{333}{\milli\tesla}$. The Hall resistance is plotted versus the amplitude $I_\mathrm{p}$ of 10-$\SI{}{\milli\second}$-long current pulses. (b) Threshold current density for switching as function of the in-plane field $\mu_\mathrm{0}H_x$.}
    \label{fig:switching}
\end{figure}

The large SOTs evidenced by the harmonic Hall effect measurements in Sect.~\ref{Sect:SOTs} can be exploited to realize current-induced magnetization switching with enhanced efficiency relative, e.g., to Pt/Co bilayers. In order to study SOT-induced switching in our multilayers with PMA, we injected 10-$\SI{}{\milli\second}$-long current pulses of variable amplitude through the Hall bar devices in the presence of an in-plane bias field $H_x$ applied parallel to the current line. This field is required to break the SOT symmetry and uniquely define the switching polarity \cite{miron2011perpendicular,manchon2019current}. Figure \hyperlink{fig:switching}{6(a)} shows representative AHE hysteresis loops as a function of current measured for Pt/Tb(2)/[Co/Tb/Co] and Pt/[Co/Tb/Co]/Tb(2). In both samples, the net magnetization switches from $+M_\mathrm{z}$ to $-M_\mathrm{z}$ when the current is parallel to the in-plane field and from $-M_\mathrm{z}$ to $+M_\mathrm{z}$ when it is antiparallel to it. The loops of Fig.~\hyperlink{fig:switching}{6(a)} have opposite polarity because the AHE has opposite sign in these two samples at room temperature [see Fig.~\hyperlink{fig:compensation}{2(c)}]. 

We repeate the measurements summarized in Fig.~\hyperlink{fig:switching}{6(a)} for all of the multilayers with PMA and several in-plane magnetic fields to characterize the threshold current density $j_\mathrm{th}$ required to switch the magnetization. The data are shown in Fig.~\hyperlink{fig:switching}{6(b)} as a function of $\mu_0 H_x$. In the macrospin approximation, the threshold switching current is given by \cite{je2018spin,lee2013threshold}
\begin{equation}\label{eq:jth}
j_{th}=\frac{2e}{\hbar}\frac{\mu_0 M_s t_M}{\xi^j_{DL}}\left(\frac{H_{k}}{2}-\frac{H_{x}}{\sqrt{2}} \right),
\end{equation}
where $t_M$ is the thickness of the magnetic layer. Although Eq.~\ref{eq:jth} does not apply to switching by domain nucleation and propagation, which is the preferred mode of magnetization reversal in mesoscopic samples \cite{baumgartner2017spatially}, one still expects $j_\mathrm{th}$ to scale with $H_{k}$ and $H_{x}$ given that these two fields determine the domain nucleation barrier. Figure~\hyperlink{fig:switching}{6(b)} shows that, despite the very large $H_{k}$ of our multilayers (see Sect.~\ref{Section:preparation}), we are able to switch all of the samples with current densities below $2.5 {\times}10^7\SI{}{\ampere\per\centi\meter^2}$ and in-plane field  $H_x \ll H_k$. This is due to the combined effect of strong dampinglike SOTs and low M\textsubscript{s} owing to the ferrimagnetic character of our multilayers. The strongest dependence of $j_\mathrm{th}$ on $H_{x}$ is found in the sample with weakest magnetic anisotropy, namely Pt/Tb(2)/[Co/Tb/Co], as expected. The samples with strongest anisotropy, where $\mu_0 H_{k} \geq 6$~T, present a much weaker dependence of $j_\mathrm{th}$ on $H_{x}$, suggesting that the in-plane field is not the main cause of reduction of the domain nucleation barrier in such a case. On the other hand, $j_\mathrm{th}$ scales with $\xi_\mathrm{DL}$ and $M_s$ in samples with comparable anisotropy, such as in the Pt/[Co/Tb/Co]/Tb($t_\mathrm{Tb}$) series. 

To investigate the role of Joule heating in the switching experiments, we measure the Hall effect by injecting different current densities up to $j \simeq 2 {\times}10^7\SI{}{\ampere\per\centi\meter^2}$. We find that the coercivity decreases by about 30\% upon increasing $j$ from $0.5$ to $2{\times}10^7\SI{}{\ampere\per\centi\meter^2}$ (not shown). Likewise, we also estimate the anisotropy field $\mu_0H_{k}$ in all the multilayers and record a drop smaller than 25\% from the case of $j<1 {\times}10^7\SI{}{\ampere\per\centi\meter^2}$ to the case of $j = 2 {\times}10^7\SI{}{\ampere\per\centi\meter^2}$.  Therefore we conclude that the Joule heating does not play a major role in determining $j_\mathrm{th}$ for $j<1{\times}10^7\SI{}{\ampere\per\centi\meter^2}$, whereas it partially assists the switching at higher current by decreasing the magnetic anisotropy energy barrier and favoring DW depinning. Indeed, $j_\mathrm{th}$ is determined by the interplay of several factors beyond $H_{k}$, $H_{x}$, $\xi_\mathrm{DL}$, and $M_s$, which include Joule heating as well as $\xi_\mathrm{FL}$, the DMI and DW pinning properties of the multilayers. 
Overall, these data show that SOTs can switch ferrimagnetic multilayers with very strong PMA at a current density of the order of $10^7\SI{}{\ampere\per\centi\meter^2}$, similarly to what is observed in CoTb alloys \cite{pham2018thermal}.

%Domain wall motion
\subsection{Current-induced domain wall motion} \label{sec:dwm}

 \begin{figure}[b]
    \centering
    \includegraphics[width=.93\textwidth]{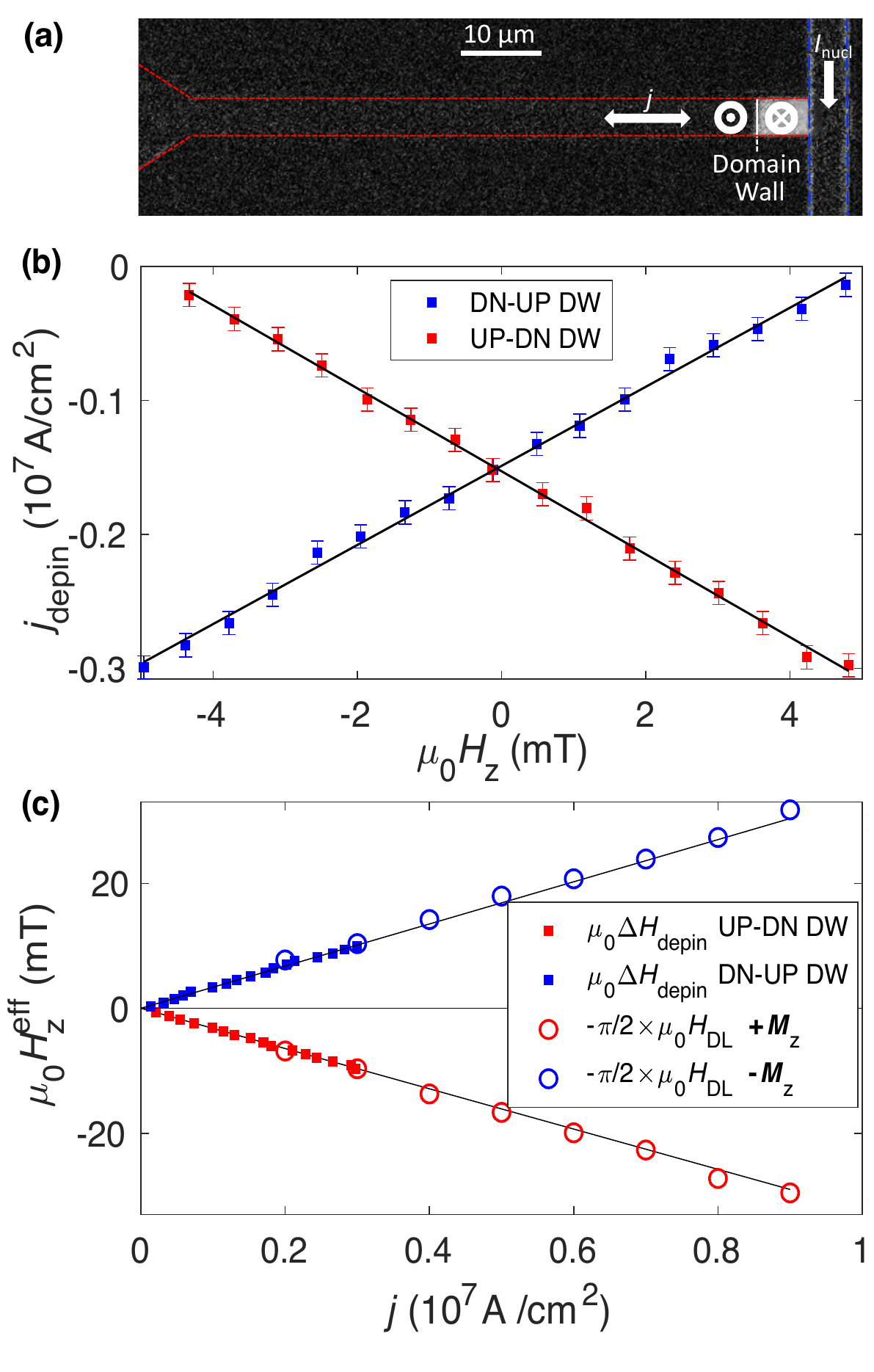}
    \caption{(a) Differential Kerr microscopy image showing the nucleation of a magnetic domain (white region) in a Pt/[Co/Tb/Co]/Tb(1) racetrack delimited by the red dotted line. (b) Critical current density required to depin up-down and down-up DWs as a function of applied out-of-plane magnetic field in Pt/[Co/Tb/Co]/Tb(1). (c) Effective out-of-plane magnetic field corresponding to the depinning field $\Delta H_\mathrm{depin}$ (squares) and dampinglike SOT effective field -$\pi/2 \times H_\mathrm{DL}$ (circles) versus current density for both DWs and magnetic states in Pt/[Co/Tb/Co]/Tb(1).}
    \label{fig:depinning}
\end{figure}

SOTs can efficiently move DWs in suitable devices and heterostructures \cite{manchon2019current}. Compensated ferrimagnets are particularly interesting for the SOT-induced DW motion because the DWs can be propelled at very high speeds with moderate currents owing to their particular dynamical properties near the magnetic and angular momentum compensation \cite{kim2017fast,caretta2018fast,siddiqui2018current,avci2019interface, cai2020ultrafast, haltz2021domain, ghosh2021current}. We investigate the current-induced DW motion in the Pt/[Co/Tb/Co]/Tb($t_\mathrm{Tb}$) samples. 
We use a wide-field Kerr microscope configured in the polar geometry to detect the out-of-plane magnetization component in  Pt/[Co/Tb/Co]/Tb($t_\mathrm{Tb}$) racetracks and study the DW response to external fields and currents. The experimental procedure can be briefly described as follows. First, the magnetization is saturated along $+z$ or $-z$ by applying an out-of-plane magnetic field larger than the coercivity. Next, a domain with the opposite magnetization is nucleated by the Oersted field produced by a short current pulse $I$\textsubscript{nuc} injected into an electrically insulated wire crossing the racetrack [Fig.~\hyperlink{fig:depinning}{7(a)}]. Finally, the DW is displaced by injecting current pulses along the DW track. We perform two types of experiments: \textit{i}) determination of the critical current for depinning a single DW in the presence or absence of an out-of-plane field ($H$\textsubscript{z}); \textit{ii}) determination of the DW velocity in the absence of any applied field.

\begin{figure}[t]
    \centering
    \includegraphics[width=.93\textwidth]{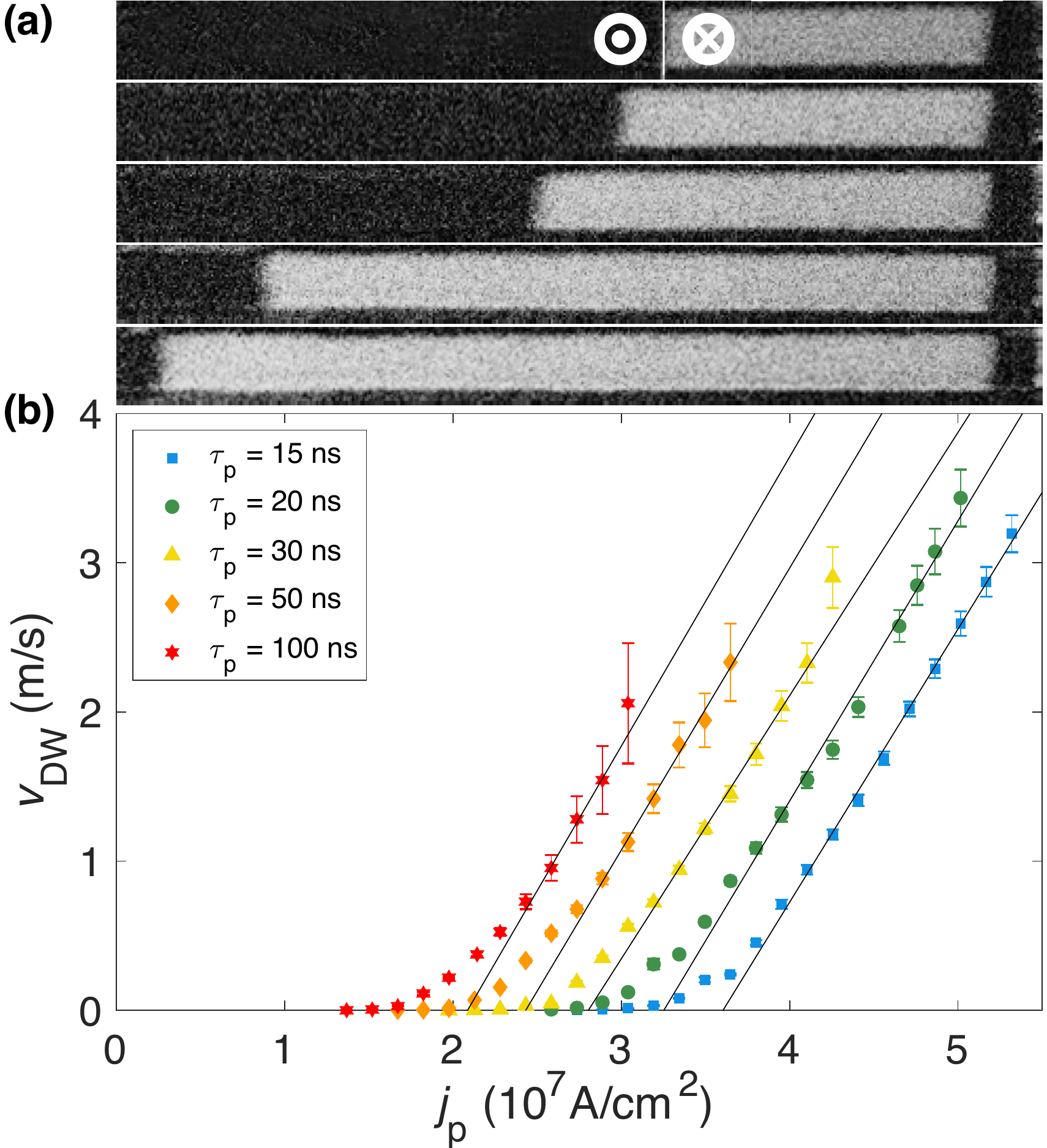}
    \caption{(a) Series of Kerr differential images showing current-driven DW motion in Pt/[Co/Tb/Co]/Tb(2) as hundreds of 20-$\SI{}{\nano\second}$-long current pulses with $j_\mathrm{p}$ = $3.6 {\times}10^7\SI{}{\ampere\per\centi\meter^2}$ amplitude are injected along the track. (b) DW velocity as a function of current density for different pulse widths $\tau_\mathrm{p} =$ 15-100 $\SI{}{\nano\second}$ in Pt/[Co/Tb/Co]/Tb(2). The errors in the DW velocity are due to the arrival time uncertainty, which is larger when a small number of pulses is used.}
    \label{fig:DWM}
\end{figure}

Figure~\hyperlink{methods}{7(b)} shows the critical current density $j$\textsubscript{depin} required to depin a DW as function of $H$\textsubscript{z} for both the up-down and down-up DWs in Pt/[Co/Tb/Co]/Tb(1). The linear variation of $j_\mathrm{depin}(H_\mathrm{z})$ shows that the current induces a dampinglike SOT that acts as an effective magnetic field along $\pm z$, assisting or hindering $H$\textsubscript{z} to depin the DWs \cite{haazen2013domain, emori2013current, emori2014spin, avci2019interface}. In the one-dimensional DW model, neglecting the spin-transfer and fieldlike torques, the effective field induced by the current is given by $H_\mathrm{z}^\mathrm{eff} = \Delta H_\mathrm{depin}$, where $\Delta H_\mathrm{depin}$ is the change in the DW depinning field under dc current injection. Alternatively, we estimate $H_\mathrm{z}^\mathrm{eff} = -\frac{\pi}{2}H_\mathrm{DL}\cos{\Psi}$, where $H_\mathrm{DL}$ is the dampinglike SOT effective field obtained by the harmonic Hall effect measurements in Sect.~\ref{Sect:SOTs} and $\Psi$ is the angle between the DW magnetization and the $x$ axis \cite{schulz2017effective}. Figure~\hyperlink{methods}{7(c)} compares $H_\mathrm{z}^\mathrm{eff}$ obtained using $\Delta H_\mathrm{depin}$ and -$\frac{\pi}{2}H_\mathrm{DL}$ for different current densities. The two measurements are in excellent agreement, which implies that the DWs are of purely homochiral Néel type (i.e., $\Psi=0\deg$). Moreover, the DW motion along the current flow indicates that they have left-handed chirality. This is consistent with the interfacial DMI occurring at the Pt/Co interface with possible additional contributions from the inner interfaces between Co and Tb and the top Tb/Ti interface \cite{kuepferling2020measuring}. Qualitatively the same results are obtained for Pt/[Co/Tb/Co]/Tb(1.5) and Pt/[Co/Tb/Co]/Tb(2), showing that the current-induced DW depinning measurements can be effectively used to identify the DW type and estimate the DL-SOT effective field in this system.
Next, we characterize the current-induced DW velocities in Pt/[Co/Tb/Co]/Tb(1, 1.5, 2). To do so, we capture a sequence of Kerr images while injecting a train of current pulses through the track to move a pre-nucleated DW, as shown in Fig.~\hyperlink{fig:DWM}{8(a)}. To estimate the DW velocity ($v\textsubscript{DW}$), we divide the total displacement by the total pulse length. Figure~\hyperlink{fig:DWM}{8(b)} shows the results from Pt/[Co/Tb/Co]/Tb(2), where $v\textsubscript{DW}$ is plotted as function of current density $j_\mathrm{p}$ for different pulse lengths $\tau_\mathrm{p}$. The data exhibit several important features. First, the current required to initiate the DW motion, i.e., the onset of the linear regime, is strongly dependent on the pulse width and is lower for longer pulses. This can be understood by the effect of thermal activation on the DW depinning, which is stronger for longer pulses. Second, in the linear regime the slope $v\textsubscript{DW}/j$ is approximately the same for all pulse widths, suggesting that Joule heating plays a minor role in $v\textsubscript{DW}$ unlike in the depinning process. In all of the cases, the depinning current in the Pt/[Co/Tb/Co]/Tb multilayers is comparable to that found in Pt/Co\textsubscript{1-x}Tb\textsubscript{x} ferrimagnetic alloys \cite{siddiqui2018current}. Third, the extrapolation of the linear fits yields velocities in the range of 10-15 m/s for $j_\mathrm{p} = 10^8\SI{}{\ampere\per\centi\meter^2}$. These values are moderate in comparison with those of other ferrimagnetic \cite{siddiqui2018current,blasing2018exchange,caretta2018fast}  and synthetic antiferromagnetic \cite{yang2015domain} systems despite the strong SOTs. However, we refrain from drawing definitive conclusions on the DW velocity behavior of our films based on the available data because we cannot conclude whether the upturn of the $v\textsubscript{DW}$ is due to the onset of a wide creep regime or the flow regime. In the former case, it is not possible to extrapolate the velocities at higher current density.  

Finally, we attempt to quantify the effective DMI field $H_\mathrm{DMI}$ in all the samples studied above. To do so, we measure $j_\mathrm{depin}$ as function of applied in-plane field up to $H_\mathrm{x}=\SI{40}{\milli\tesla}$, the maximum in-plane field available in the Kerr microscope. If $H_\mathrm{x}$ is applied opposite to $H_\mathrm{DMI}$, we expect that the DW magnetization rotates first along the $y$ axis, i.e., the DW becomes Bloch-like for $|H_\mathrm{x}| \simeq |H_\mathrm{DMI}|$, and then Néel-like with right-handed chirality for $|H_\mathrm{x}| > |H_\mathrm{DMI}|$ \cite{avci2019interface}. In the applicable in-plane field range, however, we do not observe any change in the sign of $j_\mathrm{depin}$, indicating that the chirality is not affected by $H_\mathrm{x}$ and that $|H_\mathrm{DMI}| > \SI{40}{\milli\tesla}$. These measurements show that $H_\mathrm{DMI}$ is relatively strong in Pt/[Co/Tb/Co]/Tb. 

\onecolumngrid

\begin{table}[H]

\caption{Summary of the magnetic properties and SOT efficiency of Pt/Tb($t_\mathrm{Tb}$)/[Co/Tb/Co] and Pt/[Co/Tb/Co]/Tb($t_\mathrm{Tb}$) multilayers with PMA. All values are measured in ambient conditions. Errors are a fraction of the least significant digit, except for $T_\mathrm{M}$, whose error is of the order of 1~K.}

\begin{ruledtabular}
\begin{tabular}{lccccccccccccc}
 &$\rho$ &$M_\mathrm{s}$ &$T_\mathrm{M}$
 &$R_\mathrm{AHE}$ &$R_\mathrm{PHE}$ &$\mu_\mathrm{0}H_\mathrm{k}$ &$\chi$\textsubscript{DL} &$\chi_\mathrm{FL}$ &$\xi_\mathrm{DL}^\mathrm{j}$ &$\xi_\mathrm{FL}^\mathrm{j}$&$\xi_\mathrm{DL}^\mathrm{E}$ &$\xi_\mathrm{FL}^\mathrm{E}$\\
  &($\SI{}{\micro\ohm \centi\meter}$)&($10^5$&($\SI{}{\kelvin}$)
 &($\SI{}{\milli\ohm}$) &($\SI{}{\milli\ohm}$) &($\SI{}{\tesla}$)  &($\SI{}{\milli\tesla}$ $10^{-7}$ &($\SI{}{\milli\tesla}$ $10^{-7}$ & &&($10^5$ &($10^5$\\
 
 &&$\SI{}{\ampere\per\meter}$)&&&&&$\SI{}{\ampere^{-1} \centi\meter^2})$&$\SI{}{\ampere^{-1} \centi\meter^2})$&&&$\SI{}{\Omega^{-1} m^{-1}}$) &$\SI{}{\Omega^{-1} m^{-1}}$)\\
 
\hline
Pt/Tb(1)/[Co/Tb/Co]& 72.5 & 0.24 & 310 &129
& 18 & 7.8& +72 &+23 &+0.14&+0.04 &+1.92 &+0.61 \\
Pt/Tb(1.5)/[Co/Tb/Co]& 80.3 & 0.22 & 313 &154
& 22 & 5.2 & +55 &+21 &+0.12&+0.04 &+1.44 &+0.56 \\
Pt/Tb(2)/[Co/Tb/Co]& 91.9 & 0.50 & >370 &120
& 16 & 1.1 & +17 &+7 &+0.10&+0.04 &+1.02 &+0.41\\
Pt/[Co/Tb/Co]/Tb(1)& 63.5 & 0.74 & 193 &375
& 89 & 3.2 & +22 &+20 &+0.13&+0.12 &+2.09 &+1.94\\
Pt/[Co/Tb/Co]/Tb(1.5)& 67.0 & 0.85 & 187 &425
& 101 & 3.5 & +22 &+20 &+0.18&+0.17 &+2.63 &+2.53\\
Pt/[Co/Tb/Co]/Tb(2)& 74.4 & 0.53 & 212 &341
& 62 & 2.7 & +50 &+35 &+0.29&+0.21 &+3.93 &+2.79\\

\end{tabular}
\end{ruledtabular}
\label{tab:table}

\end{table}

\twocolumngrid
  
\section{Conclusions}
Our systematic study of synthetic ferrimagnets based on Co/Tb layers shows that the magnetic properties, SOT efficiency, and switching depend strongly on the stacking sequence of Co and Tb in the heterostructures as well as on the thickness of Tb. In Pt/[Co/Tb/Co]/Tb, the dampinglike SOT efficiency reaches up to 0.3, which we attribute to the additive SOT contributions from the bottom Pt and the top Tb layer. Our conclusion is also supported by harmonic Hall measurements performed in the Tb/Co multilayers with in-plane anisotropy.  The high value of the SOT efficiency in Pt/[Co/Tb/Co]/Tb is about two and three times larger than that in Pt/Co and Pt/Tb/[Co/Tb/Co], respectively. The PMA in these multilayers is very strong with effective magnetic anisotropy fields estimated between 1 and 8~T. Despite the large PMA, the critical current for magnetization switching is relatively low, of the order of $0.5{-}2.5  {\times}10^7\SI{}{\ampere\per\meter^2}$, depending on the assisting in-plane field. Current-induced DW motion measurements reveal that the DWs are of Néel type with left-handed chirality, stabilized by interfacial DMI similar to Pt/Co bilayers.
 Our experiments show that synthetic ferrimagnetic heterostructures have comparable or improved PMA, SOT, and DMI relative to ferromagnetic systems and ferrimagnetic alloys. The stacking order offers an additional degree of freedom to tune and maximize these parameters in spintronic devices.

\section{Acknowledgments}

This work was supported by the Swiss National Science Foundation through grants no. 200020\textunderscore200465 and PZ00P2\textunderscore179944. The European Metrology Programme for Innovation and Research (EMPIR), under the Grant Agreement 17FUN08 TOPS, is also acknowledged.

\FloatBarrier

%\bibliography{bibliography}

\begin{thebibliography}{}
%\providecommand{\natexlab}[1]{#1}
%\providecommand{\url}[1]{\texttt{#1}}
%\expandafter\ifx\csname urlstyle\endcsname\relax
%  \providecommand{\doi}[1]{doi: #1}\else
%  \providecommand{\doi}{doi: \begingroup \urlstyle{rm}\Url}\fi
  
\bibitem{miron2011perpendicular}
I.~M. Miron, K.~Garello, G.~Gaudin, P.-J. Zermatten, M.~V. Costache,
  S.~Auffret, S.~Bandiera, B.~Rodmacq, A.~Schuhl, and P.~Gambardella,
  Perpendicular switching of a single ferromagnetic layer induced by in-plane current injection, Nature {\bf476}, 189 (2011).

\bibitem{liu2012spin}
L.~Liu, C.-F. Pai, Y.~Li, H.~Tseng, D.~Ralph, and R.~Buhrman, Spin-torque
  switching with the giant spin hall effect of tantalum, Science
  {\bf336}, 555 (2012).

\bibitem{garello2014ultrafast}
K.~Garello, C.~O. Avci, I.~M. Miron, M.~Baumgartner, A.~Ghosh, S.~Auffret,
  O.~Boulle, G.~Gaudin, and P.~Gambardella, Ultrafast magnetization switching by spin-orbit torques, Appl.~Phys.~Lett. {\bf105}, 212402 (2014).

\bibitem{oh2016field}
Y.-W. Oh, S.-H.~C. Baek, Y.~Kim, H.~Y. Lee, K.-D. Lee, C.-G. Yang, E.-S. Park, K.-S. Lee, K.-W. Kim, G.~Go, {\em et~al.}, Field-free switching of
  perpendicular magnetization through spin--orbit torque in antiferromagnet/ferromagnet/oxide structures, Nat.~Nanotech. {\bf11}, 878 (2016).

\bibitem{fukami2016spin}
S.~Fukami, T.~Anekawa, C.~Zhang, and H.~Ohno, A spin-orbit torque switching scheme with collinear magnetic easy axis and current configuration, Nat.~Nanotech. {\bf11}, 621 (2016).

\bibitem{avci2017current}
C.~O. Avci, A.~Quindeau, C.-F. Pai, M.~Mann, L.~Caretta, A.~S. Tang, M.~C. Onbasli, C.~A. Ross, and G.~S. Beach, Current-induced switching in a magnetic insulator,  Nat.~Mater. {\bf16}, 309 (2017).

\bibitem{grimaldi2020single}
E.~Grimaldi, V.~Krizakova, G.~Sala, F.~Yasin, S.~Couet, G.~S. Kar, K.~Garello, and P.~Gambardella, Single-shot dynamics of spin--orbit torque and spin transfer torque switching in three-terminal magnetic tunnel junctions, Nat.~Nanotech. {\bf15}, 111 (2020).

\bibitem{miron2011fast}
I.~M. Miron, T.~Moore, H.~Szambolics, L.~D. Buda-Prejbeanu, S.~Auffret,
  B.~Rodmacq, S.~Pizzini, J.~Vogel, M.~Bonfim, A.~Schuhl, {\em et~al.}, Fast current-induced domain-wall motion controlled by the rashba effect, Nat.~Mater. {\bf10}, 419 (2011).

\bibitem{emori2013current}
S.~Emori, U.~Bauer, S.-M. Ahn, E.~Martinez, and G.~S. Beach, Current-driven dynamics of chiral ferromagnetic domain walls, Nat.~Mater. {\bf12}, 611 (2013).

\bibitem{ryu2013chiral}
K.-S. Ryu, L.~Thomas, S.-H. Yang, and S.~Parkin, Chiral spin torque at
  magnetic domain walls, Nat.~Nanotech. {\bf8}, 527 (2013).

\bibitem{avci2019interface}
C.~O. Avci, E.~Rosenberg, L.~Caretta, F.~B{\"u}ttner, M.~Mann, C.~Marcus,
  D.~Bono, C.~A. Ross, and G.~S. Beach, Interface-driven chiral magnetism and current-driven domain walls in insulating magnetic garnets, Nat.~Nanotech. {\bf14}, 561 (2019).

\bibitem{luo2020current}
Z.~Luo, A.~Hrabec, T.~P. Dao, G.~Sala, S.~Finizio, J.~Feng, S.~Mayr, J.~Raabe, P.~Gambardella, and L.~J. Heyderman, Current-driven magnetic domain-wall logic, Nature {\bf579}, 214 (2020).

\bibitem{manchon2019current}
A.~Manchon, J.~\ifmmode~\check{Z}\else \v{Z}\fi{}elezn\'y, I.~M. Miron,
  T.~Jungwirth, J.~Sinova, A.~Thiaville, K.~Garello, and P.~Gambardella,
  Current-induced spin-orbit torques in ferromagnetic and antiferromagnetic systems, Rev.~Mod.~Phys. {\bf91}, 035004 (2019).

\bibitem{dieny2020opportunities}
B.~Dieny, I.~L. Prejbeanu, K.~Garello, P.~Gambardella, P.~Freitas,
  R.~Lehndorff, W.~Raberg, U.~Ebels, S.~O. Demokritov, J.~Akerman, {\em et~al.}, Opportunities and challenges for spintronics in the microelectronics industry, Nat.~Electron. {\bf3}, 446 (2020).

\bibitem{carcia1988perpendicular}
P.~Carcia, Perpendicular magnetic anisotropy in pd/co and pt/co thin-film layered structures, J.~Appl.~Phys. {\bf63}, 5066 (1988).

\bibitem{gambardella2003giant}
P.~Gambardella, S.~Rusponi, M.~Veronese, S.~Dhesi, C.~Grazioli, A.~Dallmeyer, I.~Cabria, R.~Zeller, P.~Dederichs, K.~Kern, {\em et~al.}, Giant magnetic anisotropy of single cobalt atoms and nanoparticles,  Science {\bf300}, 1130 (2003).

\bibitem{ikeda2010perpendicular}
S.~Ikeda, K.~Miura, H.~Yamamoto, K.~Mizunuma, H.~D. Gan, M.~Endo, S.~Kanai, J.~Hayakawa, F.~Matsukura, and H.~Ohno, A perpendicular-anisotropy CoFeB-MgO magnetic tunnel junction, Nat.~Mater. {\bf9}, 721 (2010).

\bibitem{dieny2017perpendicular}
B.~Dieny and M.~Chshiev, Perpendicular magnetic anisotropy at transition metal/oxide interfaces and applications, Rev.~Mod.~Phys. {\bf89}, 025008 (2017).

\bibitem{zhao2015spin}
Z.~Zhao, M.~Jamali, A.~K. Smith, and J.-P. Wang, Spin hall switching of the magnetization in ta/tbfeco structures with bulk perpendicular anisotropy, Appl.~Phys.~Lett. {\bf106}, 132404 (2015).

\bibitem{roschewsky2016spin}
N.~Roschewsky, T.~Matsumura, S.~Cheema, F.~Hellman, T.~Kato, S.~Iwata, and S.~Salahuddin, Spin-orbit torques in ferrimagnetic gdfeco alloys, Appl.~Phys.~Lett. {\bf109}, 112403 (2016).

\bibitem{ueda2016spin}
K.~Ueda, M.~Mann, C.-F. Pai, A.-J. Tan, and G.~S. Beach, Spin-orbit torques in ta/tbxco100-x ferrimagnetic alloy films with bulk perpendicular magnetic anisotropy, Appl.~Phys.~Lett.{\bf109}, 232403 (2016).

\bibitem{ueda2017temperature}
K.~Ueda, M.~Mann, P.~W. de~Brouwer, D.~Bono, and G.~S. Beach, Temperature
  dependence of spin-orbit torques across the magnetic compensation point in a ferrimagnetic tbco alloy film, Phys.~Rev.~B {\bf96}, 064410 (2017).

\bibitem{finley2016spin}
J.~Finley and L.~Liu, Spin-orbit-torque efficiency in compensated
  ferrimagnetic cobalt-terbium alloys, Phys.~Rev.~Appl. {\bf6},
  054001 (2016).

\bibitem{roschewsky2017spin}
N.~Roschewsky, C.-H. Lambert, and S.~Salahuddin, Spin-orbit torque switching of ultralarge-thickness ferrimagnetic gdfeco, Phys.~Rev.~B {\bf96}, 064406 (2017).

\bibitem{mishra2017anomalous}
R.~Mishra, J.~Yu, X.~Qiu, M.~Motapothula, T.~Venkatesan, and H.~Yang, Anomalous current-induced spin torques in ferrimagnets near compensation, Phys.~Rev.~Lett. {\bf118}, 167201 (2017).

\bibitem{seung2017temperature}
W.~Seung~Ham, S.~Kim, D.-H. Kim, K.-J. Kim, T.~Okuno, H.~Yoshikawa,
  A.~Tsukamoto, T.~Moriyama, and T.~Ono, Temperature dependence of spin-orbit effective fields in pt/gdfeco bilayers, Appl.~Phys.~Lett.  {\bf110}, 242405 (2017).

\bibitem{siddiqui2018current}
S.~A. Siddiqui, J.~Han, J.~T. Finley, C.~A. Ross, and L.~Liu, Current-induced domain wall motion in a compensated ferrimagnet, Phys.~Rev.~Lett. {\bf121}, 057701 (2018).

\bibitem{pham2018thermal}
T.~H. Pham, S.-G. Je, P.~Vallobra, T.~Fache, D.~Lacour, G.~Malinowski, M.-C. Cyrille, G.~Gaudin, O.~Boulle, M.~Hehn, {\em et~al.}, Thermal contribution to the spin-orbit torque in metallic-ferrimagnetic systems, Phys.~Rev.~Appl. {\bf9}, 064032 (2018).

\bibitem{je2018spin}
S.-G. Je, J.-C. Rojas-S{\'a}nchez, T.~H. Pham, P.~Vallobra, G.~Malinowski, D.~Lacour, T.~Fache, M.-C. Cyrille, D.-Y. Kim, S.-B. Choe, {\em et~al.}, Spin-orbit torque-induced switching in ferrimagnetic alloys: Experiments and modeling, Appl.~Phys.~Lett. {\bf112}, 062401 (2018).

\bibitem{zheng2019enhanced}
Z.~Zheng, Y.~Zhang, X.~Feng, K.~Zhang, J.~Nan, Z.~Zhang, G.~Wang, J.~Wang, N.~Lei, D.~Liu, {\em et~al.}, Enhanced spin-orbit torque and multilevel current-induced switching in w/co- tb/pt heterostructure, Phys.~Rev.~Appl. {\bf12}, 044032 (2019).

\bibitem{caretta2018fast}
L.~Caretta, M.~Mann, F.~B{\"u}ttner, K.~Ueda, B.~Pfau, C.~M. G{\"u}nther,
  P.~Hessing, A.~Churikova, C.~Klose, M.~Schneider, {\em et~al.}, Fast
  current-driven domain walls and small skyrmions in a compensated
  ferrimagnet, Nat.~Nanotech. {\bf13}, 1154 (2018).

\bibitem{cai2020ultrafast}
K.~Cai, Z.~Zhu, J.~M. Lee, R.~Mishra, L.~Ren, S.~D. Pollard, P.~He, G.~Liang, K.~L. Teo, and H.~Yang, Ultrafast and energy-efficient spin-orbit torque switching in compensated ferrimagnets, Nat.~Electron. {\bf3}, 37 (2020).

\bibitem{sala2021real}
G.~Sala, V.~Krizakova, E.~Grimaldi, C.-H. Lambert, T.~Devolder, and
  P.~Gambardella, Real-time hall-effect detection of current-induced
  magnetization dynamics in ferrimagnets, Nat.~Commun. {\bf12} 1 (2021).

\bibitem{ueda2016effect}
K.~Ueda, C.-F. Pai, A.~J. Tan, M.~Mann, and G.~S. Beach, Effect of rare earth metal on the spin-orbit torque in magnetic heterostructures, Appl.~Phys.~Lett. {\bf108}, 232405 (2016).

\bibitem{bang2016enhancement}
D.~Bang, J.~Yu, X.~Qiu, Y.~Wang, H.~Awano, A.~Manchon, and H.~Yang,
  Enhancement of spin hall effect induced torques for current-driven magnetic domain wall motion: Inner interface effect, Phys.~Rev.~B {\bf93}, 174424 (2016).

\bibitem{yu2019long}
J.~Yu, D.~Bang, R.~Mishra, R.~Ramaswamy, J.~H. Oh, H.-J. Park, Y.~Jeong,
  P.~Van~Thach, D.-K. Lee, G.~Go, {\em et~al.}, Long spin coherence length and bulk-like spin--orbit torque in ferrimagnetic multilayers, Nat.~Mater. {\bf18}, 29 (2019).

\bibitem{wong2019enhanced}
Q.~Y. Wong, C.~Murapaka, W.~C. Law, W.~L. Gan, G.~J. Lim, and W.~S. Lew,
  Enhanced spin-orbit torques in rare-earth pt/[co/ni] 2/co/tb systems, Phys.~Rev.~Appl. {\bf11}, 024057 (2019).

\bibitem{buschow1977intermetallic}
K.~Buschow, Intermetallic compounds of rare-earth and 3d transition metals, Rep.~Prog.~Phys. {\bf40}, 1179 (1977).

\bibitem{hansen1989magnetic}
P.~Hansen, C.~Clausen, G.~Much, M.~Rosenkranz, and K.~Witter, Magnetic and magneto-optical properties of rare-earth transition-metal alloys containing gd, tb, fe, co, J.~Appl.~Phys. {\bf66}, 756 (1989).

\bibitem{tanaka2010intrinsic}
T.~Tanaka and H.~Kontani, Intrinsic spin and orbital hall effects in
  heavy-fermion systems, Phys.~Rev.~B {\bf81}, 224401 (2010).

\bibitem{reynolds2017spin}
N.~Reynolds, P.~Jadaun, J.~T. Heron, C.~L. Jermain, J.~Gibbons, R.~Collette, R.~Buhrman, D.~Schlom, and D.~Ralph, Spin hall torques generated by rare-earth thin films, Phys.~Rev.~B {\bf95}, 064412 (2017).

\bibitem{wu2020spin}
Y.~C. Wu, K.~K. Meng, Q.~B. Liu, S.~Q. Zheng, J.~Miao, X.~G. Xu, and Y.~Jiang, Spin-orbit torque in antiferromagnetically coupled co and tb multilayers, Phys.~Scr. {\bf95}, 075802 (2020).
  

\bibitem{cespedes2021current}
D.~C{\'e}spedes-Berrocal, H.~Damas, S.~Petit-Watelot, D.~Maccariello, P.~Tang, A.~Arriola-C{\'o}rdova, P.~Vallobra, Y.~Xu, J.~-L.~Bello, E.~Martin, Current-Induced Spin Torques on Single GdFeCo Magnetic Layers, Adv.~Mater. {\bf33}, 2007047 (2021).


\bibitem{krishnia2021spin}
S.~Krishnia, E.~Haltz, L.~Berges, L.~Aballe, M.~Foerster, L.~Bocher, R.~Weil, A.~Thiaville, J.~Sampaio, A.~Mougin, Spin-Orbit Coupling in Single-Layer Ferrimagnets: Direct Observation of Spin-Orbit Torques and Chiral Spin Textures, Phys.~Rev.~Appl. {\bf16}, 024040 (2021).


\bibitem{thiaville2012dynamics}
A.~Thiaville, S.~Rohart, {\'E}.~Ju{\'e}, V.~Cros, and A.~Fert, Dynamics of dzyaloshinskii domain walls in ultrathin magnetic films, Europhys.~Lett. {\bf100}, 57002 (2012).

\bibitem{martinez2014current}
E.~Martinez, S.~Emori, N.~Perez, L.~Torres, and G.~S. Beach, Current-driven dynamics of dzyaloshinskii domain walls in the presence of in-plane fields: Full micromagnetic and one-dimensional analysis, J.~Appl.~Phys. {\bf115}, 213909 (2014).

\bibitem{baumgartner2017spatially}
M.~Baumgartner, K.~Garello, J.~Mendil, C.~O. Avci, E.~Grimaldi, C.~Murer,
  J.~Feng, M.~Gabureac, C.~Stamm, Y.~Acremann, {\em et~al.}, Spatially and time-resolved magnetization dynamics driven by spin--orbit torques,
  Nat.~Nanotech. {\bf12}, 980 (2017).

\bibitem{mimura1976hall}
Y.~Mimura, N.~Imamura, and Y.~Kushiro, Hall effect in
  rare-earth--transition-metal amorphous alloy films, J.~Appl.~Phys. {\bf47}, 3371 (1976).

\bibitem{avci2019effects}
C.~O. Avci, G.~S. Beach, and P.~Gambardella, Effects of transition metal spacers on spin-orbit torques, spin hall magnetoresistance, and magnetic
  anisotropy of pt/co bilayers, Phys.~Rev.~B {\bf100}, 235454 (2019).

\bibitem{webb1988coercivity}
D.~Webb, A.~Marshall, Z.~Sun, T.~Geballe, and R.~M. White, Coercivity of a macroscopic ferrimagnet near a compensation point, IEEE Trans.~Magn. {\bf24}, 588 (1988).

\bibitem{avci2014fieldlike}
C.~O. Avci, K.~Garello, C.~Nistor, S.~Godey, B.~Ballesteros, A.~Mugarza,
  A.~Barla, M.~Valvidares, E.~Pellegrin, A.~Ghosh, {\em et~al.}, Fieldlike and antidamping spin-orbit torques in as-grown and annealed ta/cofeb/mgo layers, Phys.~Rev.~B {\bf89}, 214419 (2014).

\bibitem{garello2013symmetry}
K.~Garello, I.~M. Miron, C.~O. Avci, F.~Freimuth, Y.~Mokrousov, S.~Bl{\"u}gel, S.~Auffret, O.~Boulle, G.~Gaudin, and P.~Gambardella, Symmetry and magnitude of spin--orbit torques in ferromagnetic heterostructures, Nat.~Nanotech. {\bf8}, 587 (2013).

\bibitem{ghosh2017interface}
A.~Ghosh, K.~Garello, C.~O. Avci, M.~Gabureac, and P.~Gambardella,
  Interface-enhanced spin-orbit torques and current-induced magnetization
  switching of pd/co/alo x layers, Phys.~Rev.~Appl. {\bf7}, 014004 (2017).

\bibitem{kim2013layer}
J.~Kim, J.~Sinha, M.~Hayashi, M.~Yamanouchi, S.~Fukami, T.~Suzuki, S.~Mitani, and H.~Ohno, Layer thickness dependence of the current-induced effective field vector in ta| cofeb| mgo, Nat.~Mater. {\bf12}, 240 (2013).

\bibitem{hayashi2014quantitative}
M.~Hayashi, J.~Kim, M.~Yamanouchi, and H.~Ohno, Quantitative characterization of the spin-orbit torque using harmonic hall voltage measurements, Phys.~Rev.~B {\bf89}, 144425 (2014).

\bibitem{pai2015dependence}
C.-F. Pai, Y.~Ou, L.~H. Vilela-Le{\~a}o, D.~Ralph, and R.~Buhrman, Dependence of the efficiency of spin hall torque on the transparency of pt/ferromagnetic layer interfaces, Phys.~Rev.~B {\bf92}, 064426 (2015).

\bibitem{qiu2016enhanced}
X.~Qiu, W.~Legrand, P.~He, Y.~Wu, J.~Yu, R.~Ramaswamy, A.~Manchon, and H.~Yang, Enhanced spin-orbit torque via modulation of spin current absorption, Phys.~Rev.~Lett. {\bf117}, 217206 (2016).
  
  \bibitem{avci2014interplay}
C.~O.~Avci K.~Garello, M.~Gabureac, A.~Ghosh, A.~Fuhrer, A.~F.~Santos and P.~Gambardella, Interplay of spin-orbit torque and thermoelectric effects in ferromagnet/normal-metal bilayers, Phys.~Rev.~B {\bf90}, 224427 (2014).

\bibitem{lee2013threshold}
K.-S. Lee, S.-W. Lee, B.-C. Min, and K.-J. Lee, Threshold current for
  switching of a perpendicular magnetic layer induced by spin hall effect,
  Appl.~Phys.~Lett. {\bf102}, 112410 (2013).

\bibitem{kim2017fast}
K.-J. Kim, S.~K. Kim, Y.~Hirata, S.-H. Oh, T.~Tono, D.-H. Kim, T.~Okuno, W.~S. Ham, S.~Kim, G.~Go, {\em et~al.}, Fast domain wall motion in the vicinity of the angular momentum compensation temperature of ferrimagnets, 
 Nat.~Mater. {\bf16}, 1187 (2017).

\bibitem{haltz2021domain}
E.~Haltz, S.~Krishnia, L.~Berges, A.~Mougin, and J.~Sampaio, Domain wall
  dynamics in antiferromagnetically coupled double-lattice systems, Phys.~Rev.~B {\bf103}, 014444 (2021).

\bibitem{ghosh2021current}
S.~Ghosh, T.~Komori, A.~Hallal, J.~Pe{\~n}a~Garcia, T.~Gushi, T.~Hirose,
  H.~Mitarai, H.~Okuno, J.~Vogel, M.~Chshiev, {\em et~al.}, Current-driven
  domain wall dynamics in ferrimagnetic nickel-doped mn4n films: Very large
  domain wall velocities and reversal of motion direction across the magnetic compensation point, Nano Lett. {\bf21}, 2580 (2021).

\bibitem{haazen2013domain}
P.~Haazen, E.~Mur{\`e}, J.~Franken, R.~Lavrijsen, H.~Swagten, and B.~Koopmans, Domain wall depinning governed by the spin hall effect, Nat.~Mater. {\bf12}, 299 (2013).

\bibitem{emori2014spin}
S.~Emori, E.~Martinez, K.-J. Lee, H.-W. Lee, U.~Bauer, S.-M. Ahn, P.~Agrawal, D.~C. Bono, and G.~S. Beach, Spin hall torque magnetometry of
  dzyaloshinskii domain walls, Phys.~Rev.~B {\bf90}, 184427 (2014).

\bibitem{schulz2017effective}
T.~Schulz, K.~Lee, B.~Kr{\"u}ger, R.~L. Conte, G.~V. Karnad, K.~Garcia,
  L.~Vila, B.~Ocker, D.~Ravelosona, and M.~Kl{\"a}ui, Effective field
  analysis using the full angular spin-orbit torque magnetometry dependence, Phys.~Rev.~B {\bf95}, 224409 (2017).

\bibitem{kuepferling2020measuring}
M.~Kuepferling, A.~Casiraghi, G.~Soares, G.~Durin, F.~Garcia-Sanchez, L.~Chen, C.~H. Back, C.~H. Marrows, S.~Tacchi, and G.~Carlotti, Measuring
  interfacial dzyaloshinskii-moriya interaction in ultra thin films,
  arXiv preprint arXiv:2009.11830 (2020).

\bibitem{blasing2018exchange}
R.~Bl{\"a}sing, T.~Ma, S.-H. Yang, C.~Garg, F.~K. Dejene, A.~T. N'Diaye,
  G.~Chen, K.~Liu, and S.~S.~P. Parkin, Exchange coupling torque in
  ferrimagnetic co/gd bilayer maximized near angular momentum compensation
  temperature, Nat.~Commun. {\bf9}, 4984 (2018).

\bibitem{yang2015domain}
S.-H. Yang, K.-S. Ryu, and S.~Parkin, Domain-wall velocities of up to 750 m/s driven by exchange-coupling torque in synthetic antiferromagnets, Nat.~Nanotech. {\bf10}, 221 (2015).


\end{thebibliography}
\bibliographystyle{}

\end{document}